\documentclass[showpacs,twocolumn,aps,prr,superscriptaddress,10pt]{revtex4-2}

\usepackage[usenames,dvipsnames]{xcolor}
\usepackage{amsmath}
\usepackage{amsfonts}
\usepackage{amssymb}
\usepackage{stmaryrd} 
\usepackage[T1]{fontenc}
\usepackage{mathptmx}
\usepackage[colorlinks=true,citecolor=blue,linkcolor=red]{hyperref}

\usepackage{graphicx}
\usepackage{bbold}	
\usepackage[makeroom]{cancel}		
\usepackage{multirow}				
\usepackage[normalem]{ulem}        
\usepackage{array}
\usepackage{comment}
\usepackage{pdfpages}

\newcommand{\unit}[1]{\,\mathrm{#1}} 
\newcommand{\equa}[1]{Eq.~\eqref{#1}} 
\newcommand{\fig}[1]{Fig.~\ref{#1}}

\newcommand{\rom}[1]{\uppercase\expandafter{\romannumeral #1\relax}}

\renewcommand{\vec}{\mathbf}

\usepackage{ulem}

\graphicspath{{figures/}}

\makeatletter
\AtBeginDocument{\let\LS@rot\@undefined}
\makeatother

\begin{document}
\title{Optical detection and manipulation of pseudospin orders in Wigner crystals}
	
\author{Yichen Dong}
\affiliation{State Key Laboratory of Low-Dimensional Quantum Physics and Department of Physics, Tsinghua University, Beijing 100084, People's Republic of China 
}

\author{Eugene Demler}
\email[Contact author: ]{demlere@ethz.ch}
\affiliation{
Institute for Theoretical Physics, ETH Zurich, 8093 Zurich, Switzerland 
}

\author{Zhiyuan Sun}
\email[Contact author: ]{zysun@tsinghua.edu.cn}
\affiliation{State Key Laboratory of Low-Dimensional Quantum Physics and Department of Physics, Tsinghua University, Beijing 100084, People's Republic of China 
}
\affiliation{
Frontier Science Center for Quantum Information, Beijing 100084, People's Republic of China
}


	\begin{abstract}
		In Wigner-crystal states of two-dimensional electrons, the spin ordering remains poorly understood.	
		The small energy differences between candidate spin orders make theoretical studies less reliable, and probing magnetic order at a nonzero wave vector is experimentally challenging.
		In modern realizations of Wigner crystals, the electronic spin degree of freedom is often replaced by a valley pseudospin associated with nonzero Berry curvature.
		The resulting anomalous velocity  couples the electrons' pseudospin texture  to their orbital vibration.
		We show that this mechanism enables optical detection of pseudospin orders in Wigner crystals by producing sharp
	 signatures in the terahertz optical conductivity. 
	 For example, antiferromagnetic pseudospin order enables light to excite collective electronic vibrations at the ordering wave vector, generating a characteristic absorption peak. 
		Based on the same principle, we further show that a strong optical drive generates an effective potential that reshapes the pseudospin energy landscape, inducing phase transitions  to stripe antiferromagnetic states. 
		These results point to a route for optical detection and control of spin order via its coupling to orbital motion.
	\end{abstract}

	\maketitle


\section{Introduction}	
Wigner predicted in 1934 that at low densities, electrons are driven by Coulomb interaction to form a crystal~\cite{wigner_interaction_1934,bonsall_static_1977,zhou_electronic_2026}.
Signatures of two-dimensional (2D) Wigner crystals were observed later with~\cite{Lozovik.1975} and without~\cite{Grimes.1979, Yoon.1999} magnetic fields. Recently, Wigner crystals have seen an experimental revival due to the emergence of 2D transition metal dichalcogenides (TMDCs)~\cite{smolenski_signatures_2021,zhou_bilayer_2021,Sung2025,zhou_electronic_2026,xiang_imaging_2025,chen_terahertz_2025, zhang2025transportevidencewignercrystals, zhang_wigner_2026}, graphene~\cite{tsui_direct_2024}
and their moir\'e structures~\cite{regan_mott_2020,li_imaging_2021, xu_correlated_2020, zhao_electronic_2025, qi_exciton_2026}.
Remarkably, people are now able to directly image 2D Wigner crystals with scanning tunneling microscopy~\cite{li_imaging_2021,tsui_direct_2024,xiang_imaging_2025}.

On top of the Wigner lattice, the electronic spin structure has been theoretically predicted to exhibit a rich phase diagram~\cite{Bernu.2001, drummond_phase_2009, knorzer_wigner_2020,Kim.2022, calvera_pseudo-spin_2023, Kim.2024, Esterlis.2025bilayer, Esterlis.2025} but remains experimentally mysterious~\cite{Hossain.2020,Falson2022,Sung2025}.
Moreover, the modern host materials such as 2D TMDCs~\cite{smolenski_signatures_2021,zhou_bilayer_2021,Sung2025,zarenia_wigner_2017, knorzer_wigner_2020, calvera_pseudo-spin_2023,xiang_imaging_2025}, their moir\'e bilayers~\cite{regan_mott_2020,li_imaging_2021, xu_correlated_2020, zhao_electronic_2025}, 
and rhombohedral multilayer graphene~\cite{Lu2025,
Dong.2024, 
PhysRevLett.133.206503,
Patri.2024,
PhysRevX.14.041040,
PhysRevLett.132.236601,
PhysRevB.110.205124,
PhysRevB.110.205130,
Zeng.2025,
valenti2025,
zverevich2026}
introduce valley-pseudospin degrees of freedom to the electrons/holes~\cite{xiao_coupled_2012, xu_spin_2014, manzeli_2d_2017}, 
raising intriguing questions of possible pseudospin orders~\cite{calvera_pseudo-spin_2023}
and quantum geometric effects~\cite{
Dong.2024, 
PhysRevLett.133.206503,
Patri.2024,
PhysRevX.14.041040,
PhysRevLett.132.236601,
PhysRevB.110.205124,
PhysRevB.110.205130,
Zeng.2025,
valenti2025,
zverevich2026}
 in Wigner crystals.
Unfortunately, because of the weak coupling of pseudospin orders, especially antiferromagnetic ones, to conventional transport and imaging techniques, their experimental study is limited.

In this work, we present a theoretical framework to demonstrate that the valley-pseudospin order in Wigner crystals can be both detected and manipulated using terahertz (THz) optics. 
This is because the valley pseudospin is associated with Berry curvature, so that 
it couples to the orbital motion of the electron via the anomalous velocity~\cite{Xiao.2010,xiao_coupled_2012,Dong.2025}. 
As a result, the spatial pseudospin order at momentum $q$ couples the incident light to the electronic phonons at the same momentum, which are otherwise dark.
This in turn contributes resonance peaks to the optical conductivity spectrum at the corresponding phononic frequencies. 
We further show that, because of the same reason, a strong optical driving field will generate an effective ponderomotive potential~\cite{sun_floquet_2024} that modifies the pseudospin energy landscape in the nonequilibrium regime.
The ponderomotive potential drives phase transitions from ferromagnetic into stripe antiferromagnetic (AFM) states, with the ordering wave vector tunable by the pump frequency.

\section{The model}
The phase-space Lagrangian of a Wigner crystal driven by light can be written as~\cite{Xiao.2010}
\begin{align}\label{eqn:lagrangian_WC}
L_{\text{f}} = & \sum_i  
\left(-\mathbf{k}_i \cdot \dot{\mathbf{u}}_i +  \frac{ \mathbf{k}_i^2}{2m}  
- \mathbf{A}(\mathbf{k}_i, \mathbf{S}_i) \cdot  \dot{\mathbf{k}}_i
\right)
+ \frac{1}{2}m\sum_{ij} \mathbf{u}_{i} \hat{\varphi}_{ij} \mathbf{u}_{j}
\notag \\
&-\mathbf{E} (t) \cdot \mathbf{P}[u]
\end{align}
where $\mathbf{k}_i$ and $\mathbf{u}_i$ represent the momentum and  displacement of the $i$th localized electron in the lattice and we have set the Planck constant $\hbar$ to unity.
We have focused on the case of one electron per unit cell.
The first line describes the lattice vibrations of electrons with anomalous velocity corrections, where $\mathbf{A}(\mathbf{k}_i, \mathbf{S}_i)$ is the Berry connection of the $i$th electron in momentum space (see Sec.~V A of Ref.~\cite{Xiao.2010}).
In 2D with nearly constant Berry curvature $\Omega_i=\Omega S^z_i$ associated with the pseudospin $S^z_i$, such as the valley pseudospin of low-density carriers in TMDCs~\cite{Xiao.2010,xiao_coupled_2012} and their moir\'e bilayers, one may write 
\begin{align}
\mathbf{A}(\mathbf{k}_i, \mathbf{S}_i)=\Omega_i k_i^x \hat{y}=\Omega S^z_i k_i^x \hat{y}
\,.
\end{align}
Although we are dealing with pseudospin $1/2$, the maximum values of $S^z_i$ are $\pm 1$ in our notation.
$\hat{\varphi}$ is the  long-range Coulomb potential kernel between the electronic displacements that gives rise to the spring constants for phonons.
The incident uniform light couples to matter via its dynamical electric field coupled to the total change of polarization $\mathbf{P}[u]=e\sum_i \mathbf{u}_i$. 
Note that our Lagrangian has an opposite sign to the textbook convention, so that
its static limit is the potential energy that should be minimized (instead of maximized) to obtain the mean-field-level ground/steady state.

\begin{figure}
	\centering
	\includegraphics[width=1.0\linewidth]{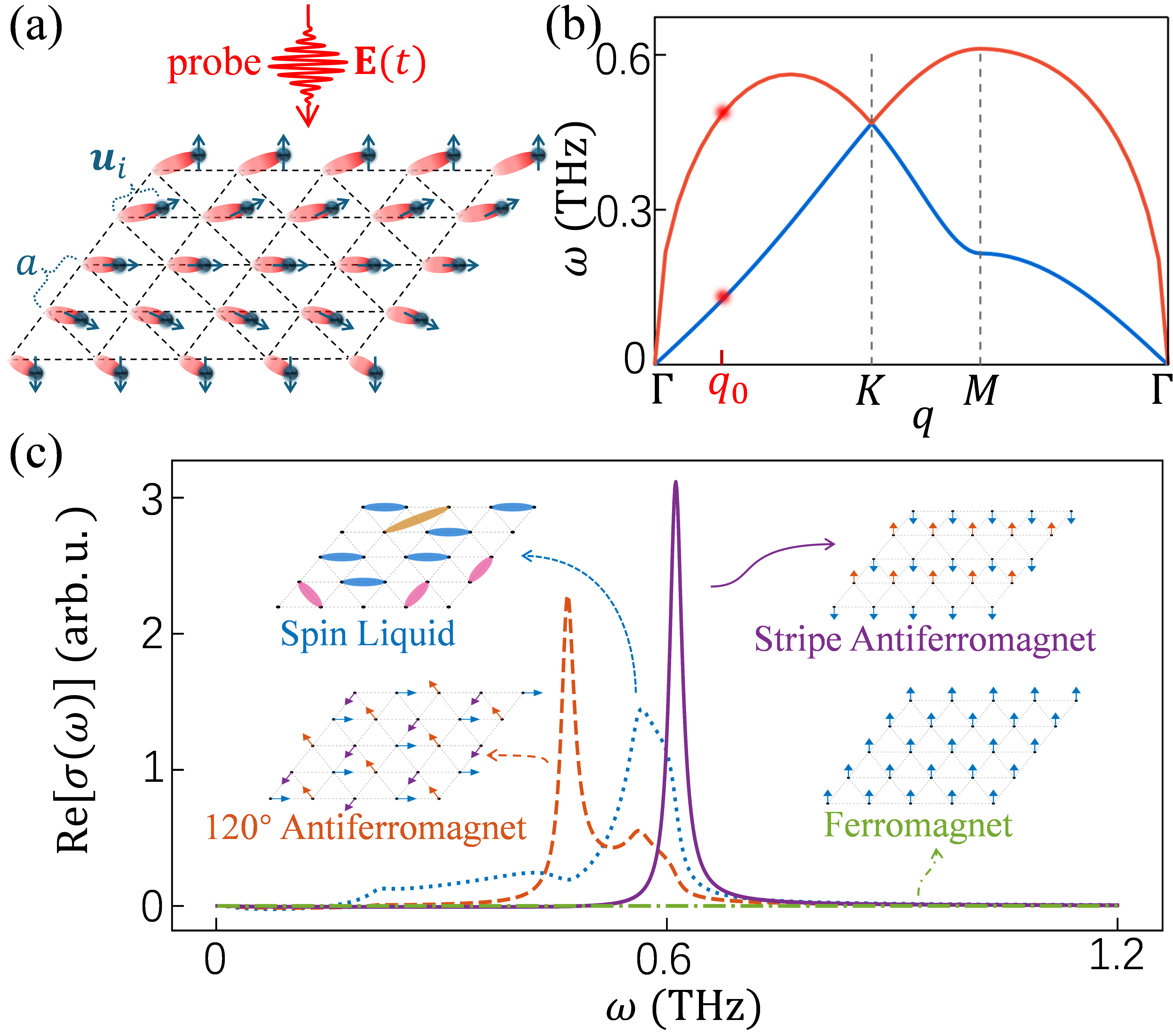}
	\caption{
	Optical detection of the pseudospin order.
		(a) Schematic of the electronic displacements in a triangular-lattice Wigner crystal driven by a dynamical uniform electric field. Because of the different anomalous velocities associated with different pseudospins, the forced motion of the electrons is not uniform, which distorts the lattice.	 
		(b) The phonon dispersion  in the limit of zero Berry curvature.
		(c) The real parts $\sigma_1(\omega)$ of the optical conductivities of a Wigner crystal  for different pseudospin configurations, where the background Drude peak is not shown. 
    The optical conductivity is shown in arbitrary units since the overall magnitude of these peaks scales sensitively with the Berry curvature, which varies tremendously across different platforms.
	We used the carrier density $n=10^{11}\unit{{cm}^{-2}}$ and the effective mass $m=m_e$ motivated by the lowest two Chern bands in twisted bilayer MoSe$_2$~\cite{Wu.2019, regan_mott_2020, deng_nonmonotonic_2025} and the phononic damping rate $\gamma=0.02 \unit{THz}$. See \fig{fig:theta_dependence} for results with physical units including pinning effects.
    }
	\label{fig:conductivity}
\end{figure}

The Hamilton's least-action principle in the phase space ($\delta S/\delta k= \delta S/\delta u=0$, where $S[k, u]=\int dt L_{\text{f}}$ is the action) yields the well-known equations of motion with anomalous velocity correction:
$
\dot{\mathbf{u}}_i = \mathbf{k}_i/m+\dot{\mathbf{k}}_i \times  (\Omega_i \hat{z})
,\,
\dot{\mathbf{k}}_i = -\partial_{\mathbf{u}_i} U,
$
where $U$ is the potential that includes the last two terms of \equa{eqn:lagrangian_WC}.
After eliminating the momenta $\mathbf{\mathit{k}}_i$ and
introducing the reduced electric field  $\mathcal{E}=e\mathbf{E}/m$ and Berry curvature $\hat{g}_i=m\Omega_i \hat{\epsilon}$, where $\hat{\epsilon}=
\begin{pmatrix}
	0 & 1 \\
	-1 & 0 
\end{pmatrix}	
$ is the antisymmetric tensor, 
the equation of motion for the lattice displacements reads
\begin{align}\label{eqn:EOM}
\ddot{\mathbf{u}}_i 
=(1- \hat{g}_i \partial_t) 
\Big(-\sum_{j} \hat{\varphi}_{ij} \mathbf{u}_j + \mathbf{\mathcal{E}}
\Big)
\,.
\end{align}
Equation (\ref{eqn:EOM}) describes the dynamics of collective electronic vibrations in the presence of a static valley-pseudospin texture. 
The latter affects the vibration dynamics via the $\hat{g}_i$ term that 
introduces valley-pseudospin-dependent anomalous velocities.
In the absence of the driving field and the Berry curvature terms, \equa{eqn:EOM} renders the bare phonon modes of the Wigner crystal whose dispersion is shown in \fig{fig:conductivity}(b).
In the following, we define $\omega_{\alpha,q}$ and 
$\Tilde{v}_{\alpha,q}=\mathbf{v}_{\alpha,q} e^{i \mathbf{q}\cdot \mathbf{R}}$
as the frequency and normal coordinate of the phonon mode of branch $\alpha$ at momentum $q$ satisfying $|\mathbf{v}_{\alpha,q}|^2=1$.
In the 2D Wigner crystal, $\alpha=\mathrm{L}$ corresponds to the longitudinal  mode [red curve in \fig{fig:conductivity}(b)] and $\alpha=\mathrm{T}$ corresponds to the transverse mode (blue curve).
In the long-wavelength limit, the longitudinal mode (red curve) is just the plasmon with the dispersion $\omega_{L,q}=\sqrt{2q\pi ne^2/m}$, where $n=2/(\sqrt{3}a^2)$ is the carrier density and $a$ is the lattice constant.



For later convenience, we quantify the pseudospin structure through the correlation function between their $z$ components
\begin{align}
\chi_{ij}=\langle S^z_i S^z_j\rangle
,\quad
\chi_q=\frac{1}{N}\sum_j e^{-i\mathbf{q} \cdot(\mathbf{R}_i-\mathbf{R}_j)} \chi_{ij}
\end{align}
where $\mathbf{R}_i$ is the spatial coordinate of the $i$th lattice site, $N$ is the total number of sites, and $\langle \cdot \cdot \cdot \rangle$ means ensemble average.
Defining the momentum-$q$ component of the $z$ pseudospins as $S_q^z=\frac{1}{N}\sum_i e^{-i\mathbf{q} \cdot\mathbf{R}_i} S_i^z$, one has $\chi_q=\langle S_{-q}^z S_q^z\rangle$,  which directly reflects the spatial ordering of $z$ pseudospins at wave vector $q$. 
A delta function structure of $\chi_q$  at momentum $q_0$ indicates long-range pseudospin order at this momentum. 
We note that if the pseudospins are along $S^x$ or $S^y$ directions, the anomalous velocity stretches the electron's wave packet instead of shifting its center of mass.
This quantum mechanical effect is not included in \equa{eqn:lagrangian_WC} and will be addressed in Sec. V.

\section{Optical Detection}
We now investigate the linear response of the Wigner crystal to a probing electric field $\mathcal{E}(t)=\mathcal{E}e^{-i\omega t}$ with \equa{eqn:EOM}. We will show that the valley-pseudospin order has characteristic signatures as peaks in the optical conductivity.
Formally, the solution can be expressed as 
\begin{align}\label{eqn:solution}
\mathbf{u}
=[\hat{G}^{-1}(\omega)+ i\omega \hat{g}  \hat{\varphi}]^{-1}
(1+i\omega\hat{g})\mathbf{\mathcal{E}}
\end{align}
where the lattice indices are absorbed in $\mathbf{u}$, $\hat{g}$ and $\hat{\varphi}$, and $\hat{G}(\omega)=(\hat{\varphi} - \omega^2)^{-1}$ is the bare retarded Green's function of the lattice to the driving field.
The Green's function has the property $\hat{G}(\omega) \Tilde{v}_{\alpha,q}=(\omega_{\alpha,q}^2-\omega^2)^{-1} \Tilde{v}_{\alpha,q}$.

In the limit of zero  Berry curvature ($\hat{g}$), the equation of motion [\equa{eqn:EOM}] exhibits Galilean invariance under uniform external field. 
In other words, all electrons are driven by the external field to move in phase with the same displacement, so that the driven dynamics is purely that of the center of mass of the whole crystal, while the internal structure of the crystal remains unchanged.
As a result, the optical response is the same as that of an ideal metal and cannot reveal the complicated single-electron band structure~\cite{brem_terahertz_2022} of a Wigner crystal unless there is pinning from disorder~\cite{fogler_dynamical_2000,Chitra.2001, Chitra2005,Dolgirev.2024} or moir\'e potential (see Appendixes for details of pinning).
Indeed, the zero $g$ limit of \equa{eqn:solution} yields $\mathbf{u}^{(0)}=-\mathcal{E}/\omega^2$ considering that the dynamical matrix $\hat{\varphi}$ satisfies $\hat{\varphi}\mathcal{E} \propto \sum_j \hat{\varphi}_{ij}=0$ because of translational invariance. This corresponds to the optical conductivity $\sigma=ine^2/(m\omega)$.
Note that translational invariance is enough to constrain the coupling of  uniform external field to the $q=0$ phonon only, so that there is only Drude response with a pole at zero frequency.
However, Galilean invariance provides a further constraint that this linear response is already exact and there are no nonlinear responses.

With the site-dependent Berry curvature $\hat{g}_i=m\Omega_i \hat{\epsilon}$ that breaks the Galilean invariance via the anomalous velocity, the motions of the electrons are no longer perfectly synchronized, as shown in \fig{fig:conductivity}(a).
As a result, the driven dynamics may distort the crystal, leading to nontrivial excitations of the phonons at nonzero momenta.
This effect must manifest as absorption peaks in the optical conductivity.
Since the anomalous-velocity effect in TMDCs is weak, we perform a perturbative expansion in $\Omega$ to resolve its impact on the optical response. 
From \equa{eqn:solution}, the first-order Berry curvature correction to the displacement is $\mathbf{u}^{(1)}
=
\hat{G}(\omega) i \omega \hat{g} \mathbf{\mathcal{E}}$. 
Its  momentum-$q$ component reads
\begin{align}\label{eqn:u1}
\mathbf{u}^{(1)}_q
=
\sum_{\alpha } 
\frac{i\omega \mathbf{v}_{\alpha,q}}{\omega_{\alpha,q}^2 - \omega^2-i\gamma \omega} 
[
\mathbf{v}_{\alpha,q} \hat{g}(q)
\mathcal{E}
]
\end{align}
where we have added a nonzero damping rate $\gamma$ for the phonons. 
The momentum summation $\sum_q$ is over the $N$ momenta in the first Brillouin zone, and goes to the dimensionless integral $\frac{N}{V_{\text{B}}}\int d^2q$ in the large system limit where the momentum volume of the first Brillouin zone is $V_{\text{B}}=8\pi^2/(\sqrt{3} a^2)$ for the triangular lattice shown in \fig{fig:conductivity}.
This correction means forced phononic oscillations at nonzero momenta $q$ and frequency $\omega$. 
They exist because the uniform driving field $\mathcal{E}$ gains a coupling to the momentum-$q$ phonon via
$\hat{g}(q)=\frac{1}{N}\sum_i e^{-i\mathbf{q}\cdot \mathbf{R}_i} \hat{g}_i
=m \Omega \hat{\epsilon} S_q^z$,  the $q$ component of the pseudospin as shown in \fig{fig:conductivity}(a), similar to disorder effect.

The zero-momentum component of $\mathbf{u}^{(1)}$ satisfies $\mathbf{\mathcal{E}} \cdot \mathbf{u}^{(1)}=0$, meaning that it is perpendicular to the applied electric field and does not contribute to the longitudinal optical conductivity $\sigma_{xx}$. 
However, the first-order lattice deformation generates an internal field via the elastic Coulomb  kernel $\hat{\varphi}$,
which drives an additional anomalous-velocity response as the second-order correction
$
\mathbf{u}^{(2)}
=-\hat{G}(\omega) i \omega \hat{g} \hat{\varphi} \hat{G}(\omega) i \omega \hat{g}
\mathcal{E}
$,
in which  the $q$ and $-q$ components from the two $\hat{g}$ operators may combine to give a nonzero uniform displacement.
Assuming the driving electric field is along $x$, the resulting second-order uniform displacement in the $x$ direction reads
\begin{align}\label{eqn:two}
u^{x(2)}_{0}
= m^2 \Omega^2 \sum_{\alpha q} 
\chi_q 
|v^y_{\alpha,q}|^2
\frac{ \omega_{\alpha,q}^2 }{\omega^2 + i\gamma \omega -\omega_{\alpha,q}^2 } 
\mathcal{E}
\,,
\end{align}
(see Appendixes for detailed derivation).
Plugging it into the current $\mathbf{J}=n e \dot{\mathbf{u}}$, one obtains the $O(\Omega^2)$ contribution to the longitudinal optical conductivity along the $x$ direction:
\begin{equation}\label{sigma}
\sigma(\omega)=ne^2 m \Omega^2 
\sum_{\alpha q} \chi_q  |v^y_{\alpha,q}|^2
\frac{i\omega \omega_{\alpha,q}^2}{\omega^2+i \gamma \omega -\omega_{\alpha,q}^2} 
\,.
\end{equation}
It is now manifest that the Berry curvature spatial disorder (such as that in a pseudospin liquid state) breaks Galilean invariance via the anomalous velocity, 
engaging phonons at nonzero momenta and introducing absorption peaks to the optical response.
Crucially, a $z$-pseudospin long-range order at nonzero momentum $q_0$, characterized by a delta-function peak in $\chi_{q}$, will generate sharp peaks in the optical absorption at the  phonon frequencies corresponding to the same $q_0$ [see \fig{fig:conductivity}(b)]. 
This connection provides a direct optical spectroscopic method for detecting pseudospin orders, and even for extracting the pseudospin structure factor $\chi_{q}$ given sufficient precision.
Note that higher-order effects of the 	Berry curvature shift the eigenfrequencies of the phonons via a self-energy.
This effect is suppressed by the small parameter $\Omega m \omega/\hbar$, which is about $10^{-3}$ in TMDC monolayers and $10^{-1}$ in their typical moir\'e bilayers~\cite{deng_nonmonotonic_2025, regan_mott_2020,li_imaging_2021, xu_correlated_2020, zhao_electronic_2025} (see Appendixes~\ref{app:berry_curvature} and \ref{app:self_energy}).
Therefore, the lowest-order result can be applied to both systems.

The optical conductivity spectra of the representative states are shown in \fig{fig:conductivity}(c).
The ferromagnetic long-range order with uniform pseudospin alignment corresponds to 
$\chi_q=\frac{V_{\text{B}}}{N}\delta(q)$ at the mean-field level, which exhibits no nonzero frequency peaks in $\sigma$ according to \equa{sigma},
consistent with Galilean invariance.
On the other hand, the stripe AFM state has $\chi_q = \frac{V_{\text{B}}}{N} \delta(q-q_0)$, leading to two sharp peaks at the phonon frequencies $\omega_{\text{L},q_0}$ and $\omega_{\text{T},q_0}$. 
There is only one peak in \fig{fig:conductivity}(c) because the transverse phonon has zero $u^y_{\text{T},q}$ for the $q_0 \parallel \hat{y}$ used there. 
We used the true long-range-ordered state for stripe AFM for later convenience. It will be shown to be stabilized by an Ising-type  interaction generated by optical driving.
For the 120$^\circ$ Néel antiferromagnetic order whose ordering momenta $q_K$ are at the $K$ points, we use $\chi_q = \frac{V_{\text{B}}}{N} \sqrt{a}|q-q_K|^{-\frac{3}{2}}/4\pi$ to capture the algebraic decay of $s_z$ fluctuations in an antiferromagnetic Heisenberg model~\cite{hu_competing_2015,jolicoeur_ground-state_1990}. 
Note that this is a result after averaging all possible orientations of the 120$^\circ$ order.
For the spin liquid and the trivial paramagnetic phases, the short-range correlation with the decay length $\xi$ is captured by a weak momentum dependence: $ \chi_q = \frac{V_{\text{B}}}{N}a^2\left[\sqrt{1/a^2+1/\xi^2}+1/\xi\right](q^2+1/\xi^2)^{-\frac{1}{2}}/2\pi$~\cite{hu_competing_2015}. 
Therefore, its optical absorption spectrum roughly reflects the phononic density of states.
Appendix~\ref{app:Distinguish} discusses how to distinguish different pseudospin orders in the optical conductivity and how to distinguish them from peaks due to other mechanisms.

\section{Optical Manipulation}
We now proceed to show that, in addition to optical detection of equilibrium pseudospin orders, the same Berry curvature effect also enables optical manipulation of pseudospin orders under strong THz  field. 
This phenomenon arises from the ponderomotive potential, the effective static potential generated  by the driving field on the low-energy degrees of freedom~\cite{sun_floquet_2024}. 
In the Wigner crystal of interest, the natural low-energy degrees of freedom are the pseudospins.
Their typical excitation energy scale is set by the ferromagnetic exchange coupling 
$J\sim \unit{\mathrm{\mu} eV}$~\cite{knorzer_wigner_2020}
which is much lower than the THz frequency of the driving field and the phonon frequency in \fig{fig:conductivity}(b).

Before the derivation, one may glean physical insights into the light-induced  potential $V_{\text{P}}(S_i)$ on the pseudospin directly from the equilibrium optical response in \equa{sigma}.
The latter shows that the pseudospin wave $\chi_q$ at momentum $q$ contributes a Lorentzian pole at frequency $\omega_{\alpha, q}$ to the optical conductivity. 
In certain cases~\cite{sun_floquet_2024}, the ponderomotive potential induced by the linearly polarized driving electrical field $2E\cos(\omega t)$ is given by $V_{\text{P}}(S_i)=E^2 \mathrm{Im}[\sigma(\omega,S_i)]/\omega$ to lowest order in the field strength.
This potential is negative for the pseudospin waves whose corresponding phonon  frequencies $\omega_{\alpha q}$ are higher than the driving frequency $\omega$. 
Therefore, the optical drive tends to push these pseudospin waves to higher values.
In other words, it will favor certain pseudospin orders,  especially the stripe AFM state at the momentum $q$ whose $\omega_{\alpha q}$ is slightly above the driving frequency.
In case the relation of the ponderomotive potential to the optical conductivity is not rigorous, we derive the former directly from the equation of motion in the following.

Formally, the total Lagrangian of the Wigner crystal is $L=L_{\text{f}}+L_{\text{s}}$, where $L_{\text{f}}$ is obtained from \equa{eqn:lagrangian_WC}. 
The pseudospin part of the lagrangian is obtained from the spin-coherent-state path integral representation of the spin Hamiltonian
$
H_0(\hat{s})
$
as
$
L_{\text{s}} =  - \sum_i  \mathcal{A}(\mathbf{S}_i) \cdot \dot{\mathbf{S}}_i 
+H_0(\mathbf{S})
$.
Here, the first term is the Berry phase term where $\nabla_s \times \mathcal{A}(\mathbf{S}) = \hat{S}$ and should be distinguished from the electronic Berry connection in \equa{eqn:lagrangian_WC}. 
The force exerted by the driven phonons on the classical spin $\mathbf{S}$ is $F_{i}=- \partial_{S_i}L_{\text{f}}=\Omega k_i^x \dot{k}_i^y$, which is purely along the $S_z$ direction [see \equa{eqn:lagrangian_WC}].
On the classical mean-field level, its time average is the ponderomotive force~\cite{sun_floquet_2024}:
\begin{align}\label{eqn:F_P}
F_{\text{P}i}&=\Omega \langle k_i^x \dot{k}_i^y \rangle_t
,\quad 
F_{\text{P}}(q)= \sum_i e^{-i\mathbf{q} \cdot\mathbf{R}_i} F_{\text{P}i}= N C_q  S_q^z
,\notag\\
C_q&= E^2 m e^2\Omega^2 \sum_{\alpha} 
|u^y_{\alpha,q}|^2
\mathrm{Re}
\left[
\frac{\omega_{\alpha q}^2}{\omega_{\alpha q}^2 - \omega^2-i\gamma\omega}
\right]
\,,
\end{align}
(see Appendix~\ref{app:ponderomotive_force} for a derivation).
Here $F_{\text{P}}(q)$ is the force felt by $\mathbf{S}_{-q}$, the momentum-$q$ component of the spin configuration, expanded to lowest order in the Berry curvature $\Omega$.
$k_i^x$ and $\dot{k}_i^y$ are found from \equa{eqn:u1} and the equation of motion relating $\mathbf{u}$ and $\mathbf{k}$.
Note that because the model in \equa{eqn:lagrangian_WC} is quadratic in the electronic displacements and momenta, there is only linear response of the phonons to the driving field but no nonlinear responses.
Therefore, the $O(E^2)$ ponderomotive force in \equa{eqn:F_P} is already an exact expansion in the driving field and there are no higher-order corrections.
The quantum and thermal fluctuation corrections to the force could be obtained by changing the time average to the  Keldysh path integral $F_{\text{P}i}=-\langle \partial_{S_i}L_{\text{f}} \rangle_{\text{path integral}}$~\cite{sun_floquet_2024,huang2025universalphasetransitionsmatter}, which is beyond the scope of this work.

Integrating the force over $\mathbf{S}_q$, one obtains the ponderomotive potential $V_{\text{P}}$ from \equa{eqn:F_P}, which renders the net effective pseudospin Hamiltonian 
\begin{equation}\label{eqn:vp}
H_{\text{eff}}=-J\sum_{\langle ij \rangle }\hat{\mathbf{s}}_i  \cdot \hat{\mathbf{s}}_j + V_{\text{P}},\quad
V_{\text{P}}=-N\sum_q C_q \hat{s}_{-q}^z \hat{s}_q^z
\,.
\end{equation}
Here, the first term is a nearest-neighbor Heisenberg ferromagnetic coupling that roughly accounts for the equilibrium exchange coupling and ring exchange effects~\cite{Bernu.2001, drummond_phase_2009, Kim.2022, Kim.2024, Esterlis.2025bilayer, Esterlis.2025} and $\langle ij \rangle$ means nearest neighbor.
The second term is the ponderomotive potential generated by the optical drive, which could also be written in real space
as $V_{\text{P}}=- \sum_{ij} J_{ij} \hat{s}_i^z \hat{s}_j^z$,
where 
$
J_{ij}= \frac{1}{N} \sum_{q} 
C_q 
e^{i \mathbf{q} \cdot (\mathbf{R_i}-\mathbf{R_j})}
$ is a nonlocal Ising type of coupling.
Note that it is exactly the formula $V_{\text{P}}(S_i)=E^2 \mathrm{Im}[\sigma(\omega,S_i)]/\omega$ predicted by the optical conductivity since $\chi_q= S_{-q}^z S_q^z$ in \equa{sigma}. 
From \equa{eqn:F_P}, $C_q$ is maximized for $\omega_{\alpha q}$ closely above $\omega$, meaning that the ponderomotive potential favors the pseudospin order with the momentum $q$ whose corresponding phonon frequency is closely above the driving frequency.

\begin{figure}
	\centering
	\includegraphics[width=1.0\linewidth]{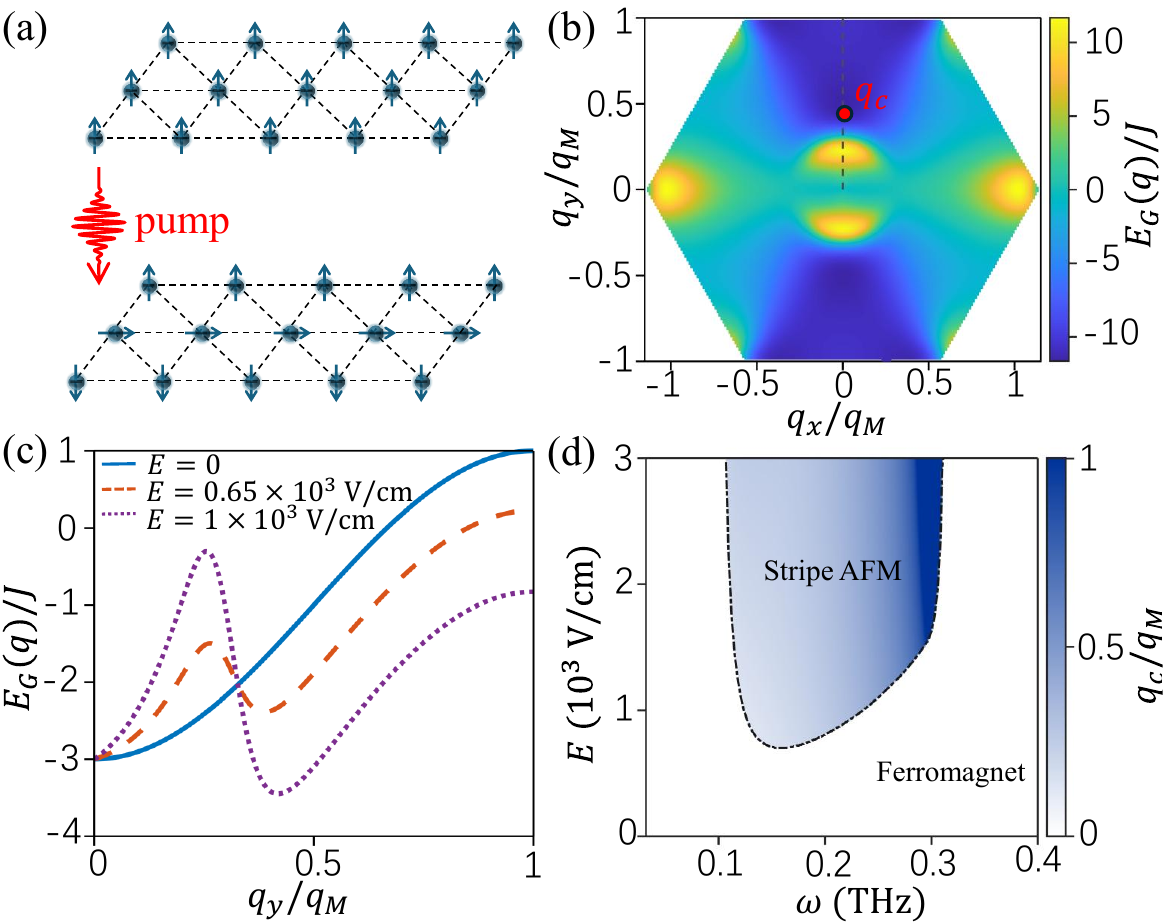}
	\caption{
    Light-manipulated pseudospin order.
		(a) Schematic illustration of the pump light switching the ferromagnetic Wigner crystal  to  a stripe antiferromagnetic state.
		 (b) The energy landscape $E_{\text{G}}(q)$ of the stripe antiferromagnetic state as a function of its ordering momentum $\mathbf{q}$ [in units of $q_M=2\pi/(\sqrt{3}a)$] under the $x$-polarized driving field $E=2.5\times10^{3} \unit{V/cm}$ at the driving frequency $\omega=0.25 \unit{THz}$. 
		 The global energy minimum corresponds to the optimized ordering momentum that is located along the $q_y$ direction. 
		 (c) Cross-sectional cuts of $E_{\text{G}}(q)$ along the gray line in panel (b) for three different pump fields, showcasing the pump induced phase transition. 
		 (d) The zero-temperature phase diagram on the plane of driving frequency $\omega$ and electric field  $E$, with the color scale encoding the pseudospin ordering wave vector $q_y$.
		 The gray dashed line marks the critical field $E_c$ required to drive the phase transition.
		 The other parameters  are $m=m_e$, the twist angle $\theta = 1.0^{\circ}$ for a twisted bilayer TMDC, filling factor $\nu = 1/10$ ($n=0.4\times10^{11}\unit{cm^{-2}}$), $\gamma=0.05 \unit{THz}$, $J=5 \unit{\mathrm{\mu} eV}$ estimated from $r_s= 134$, and the pinning frequency $\omega_0 = 0.1~\unit{THz}$ and broadening $\gamma_0 = 0.1 \unit{THz}$.
	\label{fig:phasediagram}}
\end{figure}

Next, we employ a variational approach to analyze  its phase diagram within the mean-field level. The variational wave function for the stripe AFM state with long-range valley-pseudospin order at momentum $q$ is
\begin{equation}
|q\rangle = \prod_i 
\exp\left(i \frac{\sigma_y}{2} \mathbf{q} \cdot \mathbf{R}_i \right) 
|\uparrow \rangle_i
\,
\end{equation}
where $|\uparrow \rangle_i$ means the pseudospin-up state with $s^z=1$ at site $i$ [see \fig{fig:phasediagram}(a) for an illustration].
The energy  per site   of this state is 
\begin{align}\label{eqn:Energy}
E_{\text{G}}(q)=
\frac{1}{N}\langle q|  H_{\text{eff}} |q\rangle 
=-\frac{1}{2}J\sum_{m=1}^6 \cos(\mathbf{q} \cdot \mathbf{a}_m) 
- \frac{1}{2} C_q
\end{align}
where $\mathbf{a}_m$ are the six real-space vectors pointing from a site to its six nearest neighbors on the triangular lattice.
In the absence of the driving field $E$, the energy is minimized by $q=0$ corresponding to  ferromagnetic  order. 

The optical drive induces the ponderomotive potential $-C_q$ that favors the pseudospin orders with $\omega_{\alpha q}>\omega$. 
From the Lorentzian structure of $-C_q$ in \equa{eqn:F_P}, with fixed $\alpha=\mathrm{L}$ or $\mathrm{T}$, it has a sharp minimum at $q_c$ satisfying $\omega_{\alpha q_c}-\omega \sim \gamma$.
This creates a new minimum in the total energy at roughly the same momentum, as plotted in Figs.~\ref{fig:phasediagram}(b) and \ref{fig:phasediagram}(c).
Therefore, for strong enough driving field, the new minimum becomes lower than that at zero momentum, so that the global minimum is the stripe antiferromagnetic state $|q_c\rangle$.
The phase diagram is presented in \fig{fig:phasediagram}(d)  on the plane of the driving frequency and field. 
As the driving field is tuned beyond the threshold 
$
E_c(\omega)
$, 
the system undergoes a first-order phase transition from the ferromagnetic state $|0\rangle$ into the stripe antiferromagnetic state $|q_c\rangle$.
The threshold could be estimated as 
$
E_c(\omega)\sim  
(\frac{J \gamma a^2 \omega^3 m }{\Omega^2 n^2 e^6 })^{1/2}
$ for 
$\omega_0 \ll \omega \ll W$, 
where $\omega_0$ is the pinning frequency in the pinned Wigner crystal and $W$ is the bandwidth of the electronic phonons,
and 
$
E_c(\omega)\sim  
(\frac{J \gamma}{\Omega^2 m e^2\omega})^{1/2}
$ for $\omega \approx W$.
Notably, as the driving frequency approaches $W=0.3 \unit{THz}$, the ordering wave vector $q_c=q_{M}$ reaches the $M$ point at the Brillouin zone boundary, corresponding to the stripe antiferromagnetic state in the inset of \fig{fig:conductivity}(c).

Note that the optical drive injects heat into the system by exciting the electronic phonons via the (pinned) Drude response and the pseudospin-order-induced absorption. 
From the estimation in Appendix~\ref{app:heating}, the heating may melt the Wigner crystals for the required pump fields in monolayer TMDCs.
In contrast, in TMDC moir\'e bilayers where the electronic Berry curvature $\Omega$ is much larger, the required critical pump field $E_c$ is very small and overheating can be avoided for a  wide range of twist angles.

\section{Discussion}
So far, we have focused on the center-of-mass motion of each electron at the classical level,
such that only the $z$-direction pseudospin $S^z$ affects its equation of motion from \equa{eqn:lagrangian_WC}. 
However, we note that if the pseudospin is purely along the $S^x$ or $S^y$ direction, it may be viewed as a quantum superposition of $S^z=1$ and $S^z=-1$ states.
In this case, the anomalous velocity under external field  splits the electron into two wave packets moving in opposite directions, stretching it without changing the center of mass.
This  stretching motion constitutes  a quantum mechanical correction to the optical response.
It may be roughly accounted for by a local quantum fluctuation of $z$ pseudospin as $\chi^{\text{quantum}}_{ij}=\langle S^z_i S^z_j\rangle = \delta_{ij}$ plugged into \equa{sigma}.
Therefore, this effect contributes an incoherent background to the optical absorption, which should not mask the sharp peaks of the long-range AFM orders of $z$ pseudospin or the light-induced ponderomotive potential in \equa{eqn:vp}.

In conclusion, we have theoretically predicted that pseudospin orders can be detected and manipulated in Wigner crystals using THz optics.
One promising experimental direction is  to write the stripe AFM order using a THz  pulse and read it with a second THz pulse, demonstrating a fast memory.
For Wigner crystals in non-moiré TMDCs~\cite{smolenski_signatures_2021,
zhou_bilayer_2021,Sung2025,xiang_imaging_2025,chen_terahertz_2025, zhang2025transportevidencewignercrystals, zhang_wigner_2026}, the small Berry curvature may make the optical absorption peaks in \fig{fig:conductivity}(c) too weak to detect with near-term THz spectroscopy. Complementary probes, such as Raman scattering, are nevertheless worth exploring.
Moir\'e structures made of  TMDC  and/or graphene are the best platforms for realizing our proposed effects thanks to the large Berry curvatures of their low-energy carriers (see Appendix~\ref{app:moire}).
However, for generalized Wigner crystals at high filling factors~\cite{regan_mott_2020,li_imaging_2021, xu_correlated_2020, zhao_electronic_2025} such as $\nu=1/3$,	
the continuous 	model in \equa{eqn:lagrangian_WC} is no longer accurate, and 
a quantitative prediction requires nontrivial generalization of the theory.
Beyond this prototypical example, our work also sheds light on the efficient optical control of spin/pseudospin many-body states  in generic systems via spin-orbit coupling, pushing the optical engineering of nonequilibrium matter~\cite{bao_light-induced_2022, mori_floquet_2023, rudner_band_2020} toward correlated states~\cite{Torre.2021,Murakami.2025}.

	\begin{acknowledgments}
		We thank I. Esterlis, M. M. Fogler, H. Park, and S. Zhou for helpful discussions.
		Y.D. and Z.S. acknowledge support from the National Natural Science Foundation of China (Grants No. 12421004 and No. 12374291), Beijing Natural Science Foundation (Z240005),  and the start-up grant from Tsinghua University. 
		E.D. acknowledges support from the SNSF Project No. 200021\_212899 and NCCR SPIN, a National Centre of Competence in Research, funded by the Swiss National Science Foundation (Grant No. 225153).
	\end{acknowledgments}
	
\section*{Data availability}
Numerical data for plots in this paper are available online~\cite{codesanddata}.


\appendix	

\section{Derivation of the optical conductivity}
\label{app:sigma}
In this Appendix, we expand \equa{eqn:solution} for the electronic displacement to second order in $\Omega$ which yields Eqs.~\eqref{eqn:two} and \eqref{sigma}, the Berry curvature correction to the longitudinal conductivity of the Wigner crystal.
Note that strictly speaking, $\mathbf{u}$ in \equa{eqn:solution} is a $2N$-component vector for all the displacements, while
$\hat{g}=m\Omega_1 \hat{\epsilon}_1 \oplus m\Omega_2 \hat{\epsilon}_2\oplus \cdot \cdot \cdot \oplus m\Omega_N \hat{\epsilon}_N$ and $\hat{\varphi}$ are viewed as $2N \times 2N$ matrices.
Later on, these notations are also used with the lattice indices removed. However, this should not cause a misunderstanding.

The zeroth-order term involves the contribution from the $\Gamma$-point phonon only:
\begin{align}\label{zero}
	\mathbf{u}^{(0)} = \hat{G}(\omega) \mathbf{\mathcal{E}}= -\mathbf{\mathcal{E}}/\omega^2
	\,.
\end{align}
Consequently, the optical conductivity corresponding to this term exhibits a Drude conductivity.

The first-order term in \equa{eqn:u1} is straightforward to derive from expanding \equa{eqn:solution}.
There are two terms that may contribute: The anomalous velocity response to the external field, and that to the internal electric field generated by zeroth-order motion. However, the latter is zero because $\hat{\varphi}\mathcal{E}=0$, i.e., the zeroth-order motion corresponds to uniform displacement of the lattice and induces no internal field. The former term reads  
\begin{equation}\label{one}
	\begin{split}
		\mathbf{u}^{(1)}=\hat{G}(\omega) i\omega\hat{g}\mathbf{\mathcal{E}} = \sum_{\alpha q} 
		\frac{i\omega}{\omega_{\alpha,q}^2-\omega^2} \Tilde{v}_{\alpha,q} \Tilde{v}_{\alpha,q}^{\mathrm{T}} \hat{g}\mathbf{\mathcal{E}} 
	\end{split}
\end{equation}
where we inserted the complete basis $\hat{I}=\sum_{\alpha q}\Tilde{v}_{\alpha,q} \Tilde{v}_{\alpha,q}^{\mathrm{T}}$.
Writing out the lattice index explicitly and focusing on the Fourier component at $q$, one obtains \equa{eqn:u1}.

The second-order term arises from the anomalous velocity generated from the lattice potential induced by the first-order deformations. 
Expanding \equa{eqn:solution} to $O(\hat{g}^2)$ and making use of $\hat{\varphi}\mathcal{E}=0$ again, one obtains
\begin{align}\label{twoex}
	\mathbf{u}^{(2)}&=
	-\hat{G}(\omega) i \omega \hat{g} \hat{\varphi} \hat{G}(\omega) i \omega \hat{g} \mathbf{\mathcal{E}} 
	\notag\\
	&= \omega^2 
	\sum_{\beta q', \alpha q} 
	\frac{
		\Tilde{v}_{\beta,q'} \Tilde{v}_{\beta,q'}^{\mathrm{T}}}{\omega_{\beta, q'}^2 - \omega^2} \hat{g} \omega_{\alpha, q}^2 
	\frac{\Tilde{v}_{\alpha,q} \Tilde{v}_{\alpha,q}^T}{\omega_{\alpha, q}^2 - \omega^2} 
	\hat{g}  
	\mathcal{E} 
	\,.
\end{align}
Selecting the $q'=0$ component, one obtains the zero-momentum displacement
\begin{align}\label{eqn:two_appendix}
	\mathbf{u}^{(2)}_{0}
	=  \sum_{\alpha \beta q} 
	\frac{m^2 \Omega^2 \chi_q \omega_{\alpha,q}^2 }{\omega^2 + i\gamma \omega -\omega_{\alpha,q}^2 } 
	\mathbf{v}_{\beta,0} 
	\left(\mathbf{v}_{\beta,0} \hat{\epsilon} \mathbf{v}_{\alpha,q} \right)
	\left(\mathbf{v}_{\alpha,q} \hat{\epsilon} \mathcal{E}\right)
	\,
\end{align}
where a phenomenological damping rate $\gamma$ has been added.
Focusing on the driving field along $x$, one obtains \equa{eqn:two} and
\equa{sigma}.
\\


\section{Derivation of the ponderomotive force}
\label{app:ponderomotive_force}
In this Appendix, we derive the expression for the ponderomotive force acting on $\mathbf{S}_q$  given by \equa{eqn:F_P}. The ponderomotive force experienced by the $i$th electron is given by $F_{\text{P}i} =\Omega \langle k_i^x \dot{k}_i^y \rangle_t$, where $k_i^x$ and $\dot{k}_i^y$ could be found from the Hamilton equation of motion
\begin{align}\label{eq:k}
	k_i^x = \dot{u}_i^x - m\Omega \dot{k}_i
	,\quad
	\dot{k}_i^y = -m \left[ \hat{\varphi}\mathbf{u}\right]_i^y
	\,
\end{align}
for the driving field along the $x$ direction.
The displacement $\mathbf{u}$ could be found from Eqs. \eqref{zero} and \eqref{one}.
Because $\mathbf{u}^{(0)}$ is the uniform motion of electron lattice, it satisfies $\hat{\varphi}\mathbf{u}^{(0)}=0$. 
Thus the leading-order result for $k^y(q)=\frac{1}{N}\sum_i e^{-i\mathbf{q} \cdot \mathbf{R}_i} k^y_i $ comes from the $\mathbf{u}^{(1)}$ term in \equa{eqn:u1}:
\begin{align}\label{ky}
	-i\omega k^y(q,\omega) 
	=
	\sum_{\alpha } 
	\frac{-i\omega \omega_{\alpha,q}^2 v_{\alpha,q}^y}{\omega_{\alpha,q}^2 - \omega^2-i\gamma \omega} 
	[
	\mathbf{v}_{\alpha,q} \hat{g}(q)
	\mathcal{E}
	]
	\,.
\end{align}
which is $O(\Omega)$. 
Therefore, for the $O(\Omega^2)$ result of the ponderomotive force, one just needs the zeroth-order result of $k_i^x$:
$
k_i^x(\omega)  = ieE/\omega
$.
Plugging them in the expression $F_{\text{P}i} =\Omega \langle k_i^x \dot{k}_i^y \rangle_t$
and taking the combination between positive and negative frequency components, one obtains the ponderomotive force for the Fourier component $\mathbf{S}_{-q}$ in \equa{eqn:F_P}.
\\

\section{Berry curvature in TMDCs and their  moir\'e bilayers}
\label{app:berry_curvature}
In monolayer TMDCs, the effective two-band Hamiltonian for low-energy electrons could be written as~\cite{xiao_coupled_2012,Xiao.2010, xiao_valley-contrasting_2007}
$
H = v(\tau k_x\hat{\sigma}_x + k_y \hat{\sigma}_y) + \Delta \hat{\sigma}_z
$,
where $v$, $\Delta=mv^2$, and $\tau=\pm 1$ are the asymptotic velocity, the mass, and the valley index, respectively. 
The Berry curvature $\Omega(\mathbf{k})= \mathbf{\nabla}_{\mathbf{k}} \times \langle u_n(\mathbf{k}) | i \nabla_{\mathbf{k}} | u_n(\mathbf{k}) \rangle$ of the electrons in the conduction band is~\cite{xiao_valley-contrasting_2007, Xiao.2010}
\begin{equation}
	\Omega(\mathbf{k}) 
	= -\tau \frac{v^2 \Delta}{2\left[\Delta^2 + v^2 k^2\right]^{3/2}}\hat{z}
	\xrightarrow{k \ll mv}
	-\tau \frac{v^2}{2\Delta^2}
\end{equation}
and that for the valence band is opposite.
At low densities, the typical electronic momentum in a Wigner crystal is much smaller than $mv$, so that the Berry curvature is approximated by the constant value in the main text. 
Given the typical parameters in TMDCs, $\Omega$ falls in the range of $10-20 \unit{\mathring{\mathrm{A}}^2}$~\cite{xiao_coupled_2012}. 
To obtain a sense of its magnitude, in an electric field of $E=10^3 \unit{V/cm}$ with a driving frequency of $\omega=0.5 \unit{THz}$, the normal velocity of the forced motion of a single electron is $v_n=3.5 \times 10^4 \unit{m/s}$. 
The ratio between its anomalous velocity and the normal velocity is $v_a/v_n = \Omega m \omega/\hbar \sim 10^{-3}$.

In TMDC moir\'e bilayers~\cite{regan_mott_2020,li_imaging_2021, xu_correlated_2020, zhao_electronic_2025}  at small twist angles, the lowest two electronic bands are Chern bands with Chern numbers $C=1$ and $C=-1$ corresponding to the two valleys~\cite{Wu.2019}. 
Therefore, the Chern number is  also the pseudospin index.
The Berry curvature of the electrons in these bands may be estimated as $\Omega \approx  C/S_{\mathrm{MBZ}}$, where  $S_{\mathrm{MBZ}}$ is the area of the moir\'e Brillouin zone,  meaning that $\Omega \propto 1/\theta^2$ for twisted TMDC homobilayers with the twist angle $\theta$.
Because of the much smaller size of the moir\'e Brillouin zone for typical moir\'e bilayers~\cite{deng_nonmonotonic_2025, regan_mott_2020,li_imaging_2021, xu_correlated_2020, zhao_electronic_2025}, the Berry curvature (with the typical moir\'e period of $20 \unit{nm}$)  is $\sim 10^2$ times   larger than that in monolayer TMDCs so that $v_a/v_n \sim 10^{-1}$.
\\

\begin{figure}
	\centering
	\includegraphics[width=0.8\linewidth]{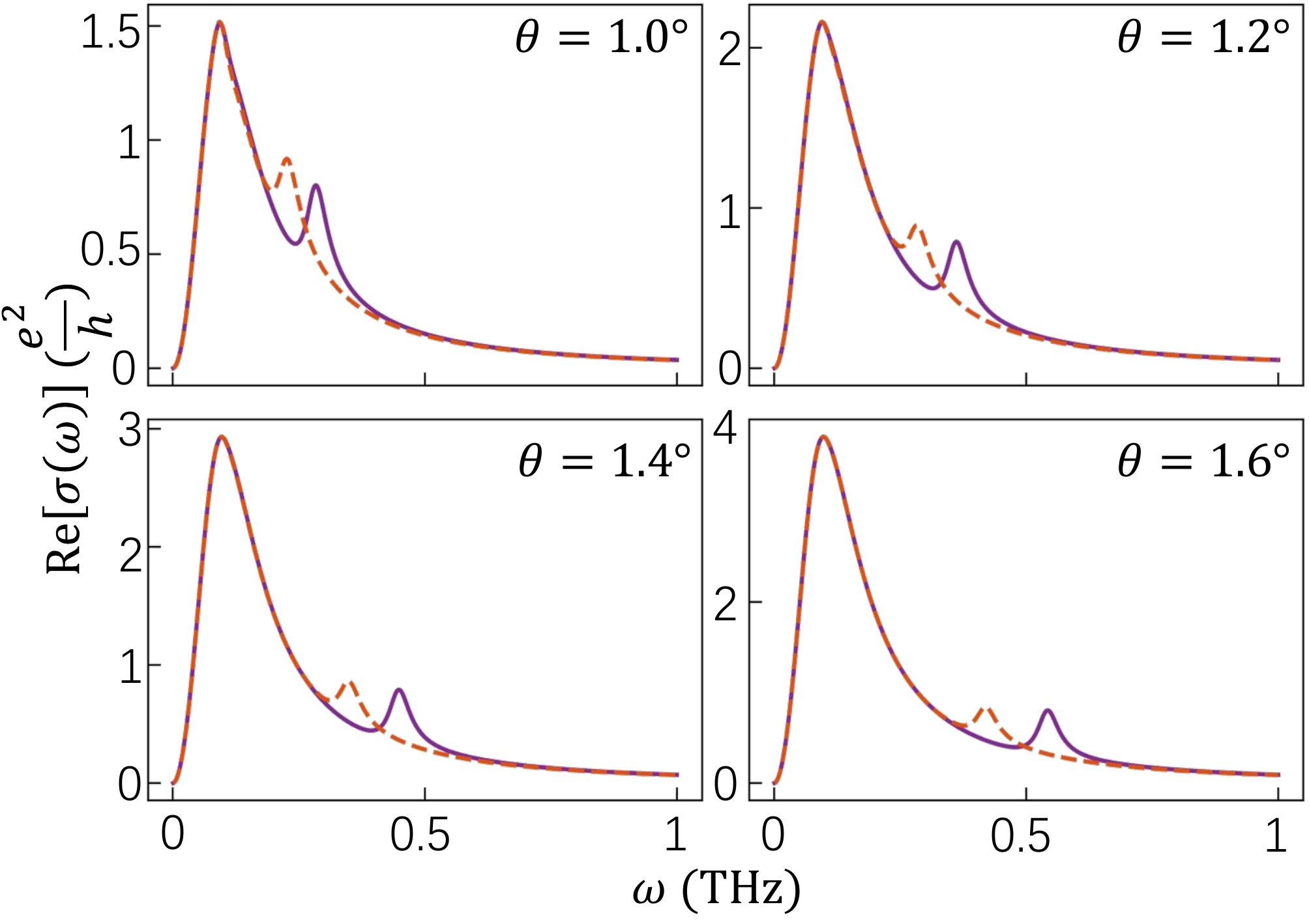}
	\caption{Twist angle ($\theta$) dependence of the optical conductivity of the (generalized) Wigner crystals in a twisted bilayer TMDC. Results are shown for four twist angles ranging from $1.0^\circ$ to $1.6^\circ$ at the same filling factor $\nu = 1/10$. 
		The orange dashed curves and purple solid curves are for Wigner crystals with the $120^\circ$ AFM order and stripe AFM order, respectively. 
		The peak close to zero frequency is the pinned Drude peak and that at higher-energy is the pseudospin-order-induced peak.
		We used the effective mass $m=m_e$~\cite{deng_nonmonotonic_2025}, the  frequency $\omega_0=0.1 \unit{THz}$ and broadening $\gamma_0=0.15 \unit{THz}$ for the pinning mode~\cite{chen_terahertz_2025}, and the damping rate $\gamma=0.05 \unit{THz}$~\cite{zhao_electronic_2025} for electronic phonons.
		\label{fig:theta_dependence}}
\end{figure}

\section{Effect of pinning}
\label{app:pinning}
Disorder potential pins the Wigner crystal and brings a gap and broadening to the phonons~\cite{fogler_dynamical_2000,Chitra.2001, Chitra2005,Dolgirev.2024}.
Moreover, moir\'e superlattices are found to host generalized Wigner crystals whose period is commensurate with the underlying moir\'e lattice~\cite{regan_mott_2020,li_imaging_2021}.
There the phonons are naturally gapped because the Wigner crystal is pinned by the moir\'e lattice even without disorder.
The simplest way to model the pinning effect is to add a harmonic potential term
$
V_{\text{pin}}
=\sum_{i}  m \omega_0^2 \mathbf{u}_{i}^2/2
$
to the Lagrangian in \equa{eqn:lagrangian_WC},
which could also be absorbed in the $\varphi$ kernel. 
As a result, the phonons obtain a gap equal to the pinning gap $\omega_0$, while their normal coordinates $\tilde{v}_{\alpha,q}$ remain unchanged since the pinning potential matrix is proportional to identity.
In the limit of zero Berry curvature, the Drude pole of the optical conductivity is shifted to $\omega_0$ as revealed by the Green's function at zero momentum: 
$\hat{G}\Tilde{v}_{\alpha,0}=(\omega_{0}^2-\omega^2)^{-1} \Tilde{v}_{\alpha,0}$. 
The zeroth-order optical conductivity is thus 
\begin{align}\label{eqn:zero_appendix_pinning}
\sigma=\frac{ine^2}{m}\frac{\omega}{\omega^2+i\gamma_0	\omega-\omega_{0}^2}
	\,
\end{align}
where the phenomenological damping rate $\gamma_0$ is explicitly added.
Furthermore, due to the breaking of translational symmetry, the first-order term $\propto (\varphi + \omega^2_0) \hat{G}(\omega) \mathbf{\mathcal{E}}$ from Eq.~\eqref{eqn:solution} is no longer zero, and the total first-order displacement equation \eqref{one} is modified by a factor and an overall shift of the poles:
\begin{equation}\label{one_pinning}
	\begin{split}
		\mathbf{u}^{(1)}  =
        \frac{\omega^2}{\omega^2-\omega_0^2}
        \sum_{\alpha q} 
		\frac{i\omega}{\omega_{\alpha,q}^2+\omega_0^2-\omega^2} \Tilde{v}_{\alpha,q} \Tilde{v}_{\alpha,q}^{\mathrm{T}} \hat{g}\mathbf{\mathcal{E}} 
	\end{split}
\,.
\end{equation}
As a result, Eqs.~(\ref{eqn:two}) and (\ref{sigma}) are modified to
\begin{align}\label{eqn:two_appendix_pinning}
	u^{x(2)}_{0}
	= 
	 \frac{\omega^4 m^2 \Omega^2}{(\omega^2+i\gamma_0 \omega-\omega_0^2)^2}
	\sum_{\alpha q} 
	\frac{ \chi_q |v^y_{\alpha,q}|^2 (\omega_{\alpha,q}^2 + \omega_0^2)  }{\omega^2 + i\gamma \omega -\omega_{\alpha,q}^2 -\omega_0^2} 
	\mathcal{E}
	\,
\end{align}
and 
\begin{equation}\label{sigma_appendix_pinning}
	\sigma(\omega)=
	 \frac{\omega^4 ne^2 m \Omega^2}{(\omega^2+i\gamma_0 \omega-\omega_0^2)^2}
	\sum_{\alpha q} 
	\frac{\chi_q  |v^y_{\alpha,q}|^2 i\omega (\omega_{\alpha,q}^2 + \omega_0^2)}{\omega^2+i \gamma \omega -\omega_{\alpha,q}^2 - \omega_0^2} 
	\,.
\end{equation}
Physically,  every phonon mode acquires a frequency shift due to the pinning, resulting in the dispersion $\omega_{\alpha,q}'=\sqrt{\omega_{\alpha,q}^2+\omega_0^2}$.
Nevertheless, the qualitative conclusions are unchanged: The pseudospin order leads to sharp absorption peaks at the renormalized phonon frequencies $\omega_{\alpha,q}'$ corresponding to the ordering wave vector $q_0$. 
Their spectral weights are modified by a factor of
$(\omega_{\alpha,q_0}'/\omega_{\alpha,q_0})^6$
, which is $O(1)$ in the weak-pinning case.
Similarly, the light-induced ponderomotive potential in \equa{eqn:F_P} is modified by the  factor $\frac{\omega_2}{\omega^2-\omega_0^2}$ and the same shift of phonon frequencies.
\\

\section{THz optical response in moir\'e bilayers}
\label{app:moire}
In this Appendix, we compute the THz optical conductivities of Wigner crystals in TMDC moir\'e bilayers.
We show that the pseudospin-order-induced peaks depend  systematically on the twist angle $\theta$ and carrier density $n$.

Figure \ref{fig:theta_dependence} shows the evolution of the  optical conductivity with twist angle $\theta$ for a homobilayer at a fixed filling factor $\nu = 1/10$. 
For $\theta = 1.0^\circ$, the large moir\'e period $L_m = 17~\unit{nm}$ corresponds to a large Berry curvature $\Omega = 8 \times 10^2\unit{\mathring{A}^2}$.
The pseudospin-order-induced peak is pronounced and remains resolvable even in the presence of a Drude background (the pinning peak). 
As the twist angle increases, pseudospin-order-induced peaks become weaker mainly because the Berry curvature decreases.

\begin{figure}
	\centering
	\includegraphics[width=0.8\linewidth]{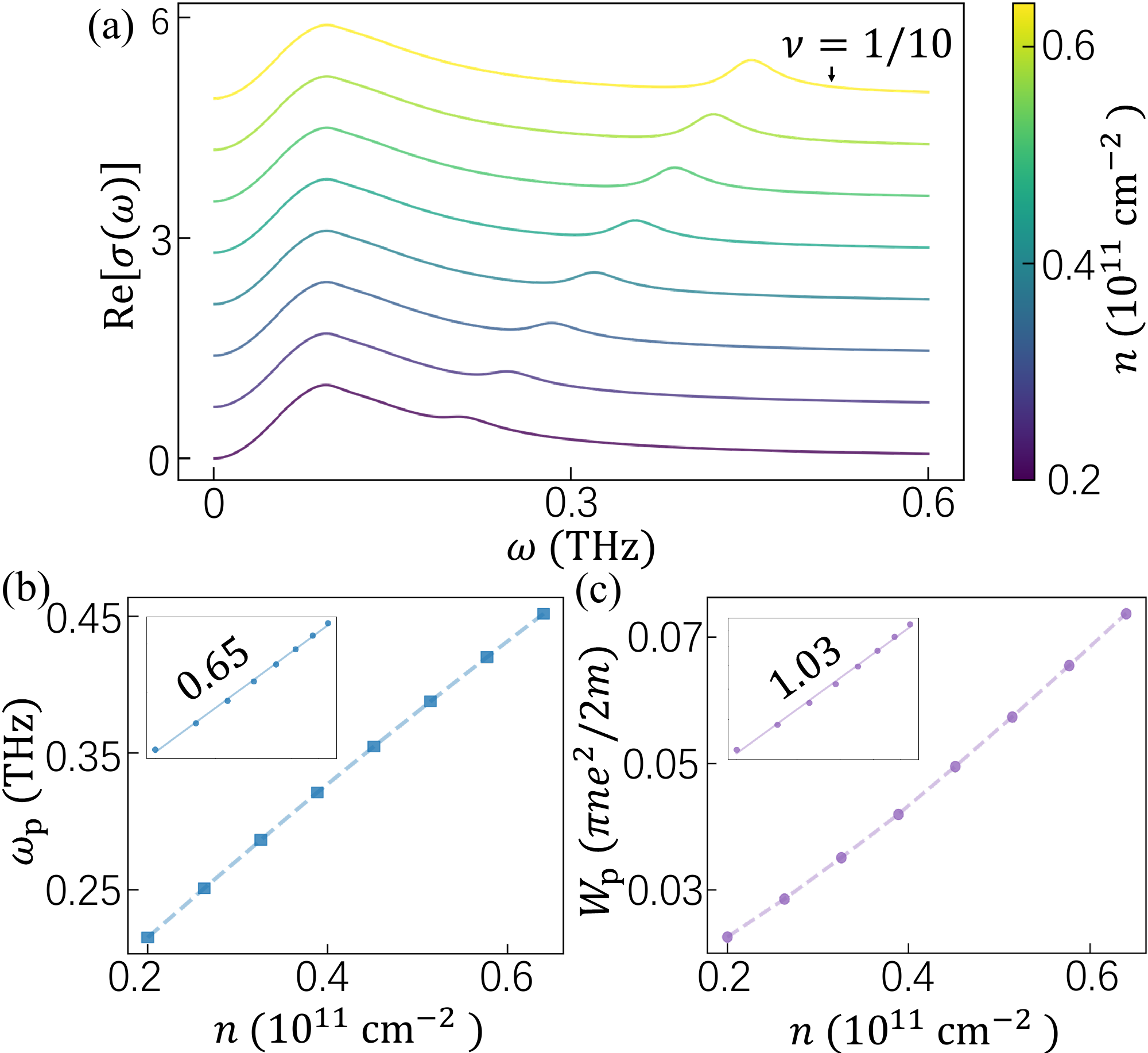}
	\caption{Carrier density ($n$) dependence of the optical response of (generalized) Wigner crystals with stripe AFM pseudospin order [see inset of \fig{fig:conductivity}(c)] in a twisted bilayer TMDC  at the twisted angle $\theta=1.4^{\circ}$. 
	The other parameters are the same as those in  \fig{fig:theta_dependence}.
    (a) 
    Different curves are the optical conductivities corresponding to different carrier densities indicated by the color bar. 
    All curves are normalized by the Drude peak and shifted vertically for clarity. 
    The maximal filling factors $\nu = 1/10$ is labeled.
    The low-energy peak is the pinned Drude peak and the high-energy one is that from pseudospin order. 
    (b) and (c) Dependence of the frequency $\omega_p$ and spectral weight $W_p$ of the pseudospin-order-induced peak on the carrier density $n$. 
    Inset: Log-log plot of $\omega_p$ and $W_p$ vs $n$, 
    with the fitted powers labeled there. 
	\label{fig:density_dependence}}
\end{figure}

Figure \ref{fig:density_dependence} demonstrates the carrier density dependence of the optical response for stripe AFM pseudospin order at the twist angle $\theta=1.4^{\circ}$. Without pinning, as the carrier density $n$ increases, the  pseudospin-order-induced peak frequency increases as $\omega_p =\omega_{q_0\alpha}\propto n^{3/4}$ and its spectral weight increases as $W_p \propto \omega_p^2 \propto n^{3/2}$ [see \equa{sigma}].
A weak pinning with the pinning frequency $\omega_0$ modifies the ideal power-law dependence by introducing a small offset:
$\omega_p \approx \omega_{q_0\alpha}+\frac{1}{2}\frac{\omega_0^2}{\omega_{q_0\alpha}}$,
$W_p = \omega_{q_0\alpha}^2+\omega_0^2$, 
 slightly modifying scaling exponents as shown in Figs.~\ref{fig:density_dependence}(b) and \ref{fig:density_dependence}(c). 
This sensitive dependence on $n$ may distinguish the pseudospin-order-induced peak from other peaks.
For example, the peaks from optical phonons in the host material (or the substrate) vary only weakly with the carrier density.
The location of the Drude peak pinned by disorder scales as $\omega_{\text{pinning}}\propto n^{3/8}$~\cite{chen_terahertz_2025} and its spectral weight scales as $W_{\text{pinning}}\propto n$, completely different power laws.
Furthermore, the (pinned) Drude peak is always there and is easy to identify. 
If the THz probe covers both the pinned Drude peak and the pseudospin-order-induced peak, the latter is most likely the higher-energy one.
\\

\section{Joule heating effect}
\label{app:heating}
When the Wigner crystal is driven by a strong laser pulse, the energy absorption by  the Drude component of the optical conductivity (or the pinning mode in the pinned case) leads to Joule heating.
In this Appendix, we estimate the heating effect characterized by an effective temperature $T_{\mathrm{eff}}$ after pumping, and identify the threshold for melting the Wigner crystal.
From the estimation, the heating is severe for the required pump fields in Wigner crystals realized in monolayer TMDC.
To avoid melting the Wigner crystal, one could either use very clean devices so that the width of the pinning mode is much smaller than our used value ($\gamma_0=0.1 \unit{THz}$), or use a substrate that could take away the heat quickly.
In contrast, in TMDC moir\'e bilayers where the electronic Berry curvature $\Omega$ is much larger, the required critical pump field $E_c$ is very small and overheating can be easily avoided.

The most promising experimental realization of writing the stripe AFM order  would be using an ultrafast THz pump pulse.
Due to the transient nature of the state induced by the ultrafast pump pulse, the injected energy is not distributed in the equilibrium way, making it challenging to determine whether the Wigner crystal melts.
Note that although most of the heating comes from driving the pinned (``zero'') mode at zero momentum, the absorbed energy is actually stored in the electronic phonon modes at nonzero momenta whose mode frequency is equal to the pump frequency $\omega$.
This is because the microscopic mechanism of optical absorption (apparently from a phenomenological damping $\gamma_0$) is the resonant excitation of the frequency-$\omega$ modes whose optical matrix elements come from momentum relaxing scattering by, e.g., disorder potential.
Therefore, the resulting phonon distribution is out of equilibrium and the temperature is not well defined. 
Since the nonlinear interactions between phonons are in general weak, we do not expect thermalization between phonons to occur within our time of interest.
In the following, we calculate the mean fluctuation of the electronic lattice due to the non-thermal distribution of electronic phonons, and compare it to the threshold fluctuation that melts the Wigner crystal in equilibrium.

\begin{figure}
	\centering
	\includegraphics[width=0.9\linewidth]{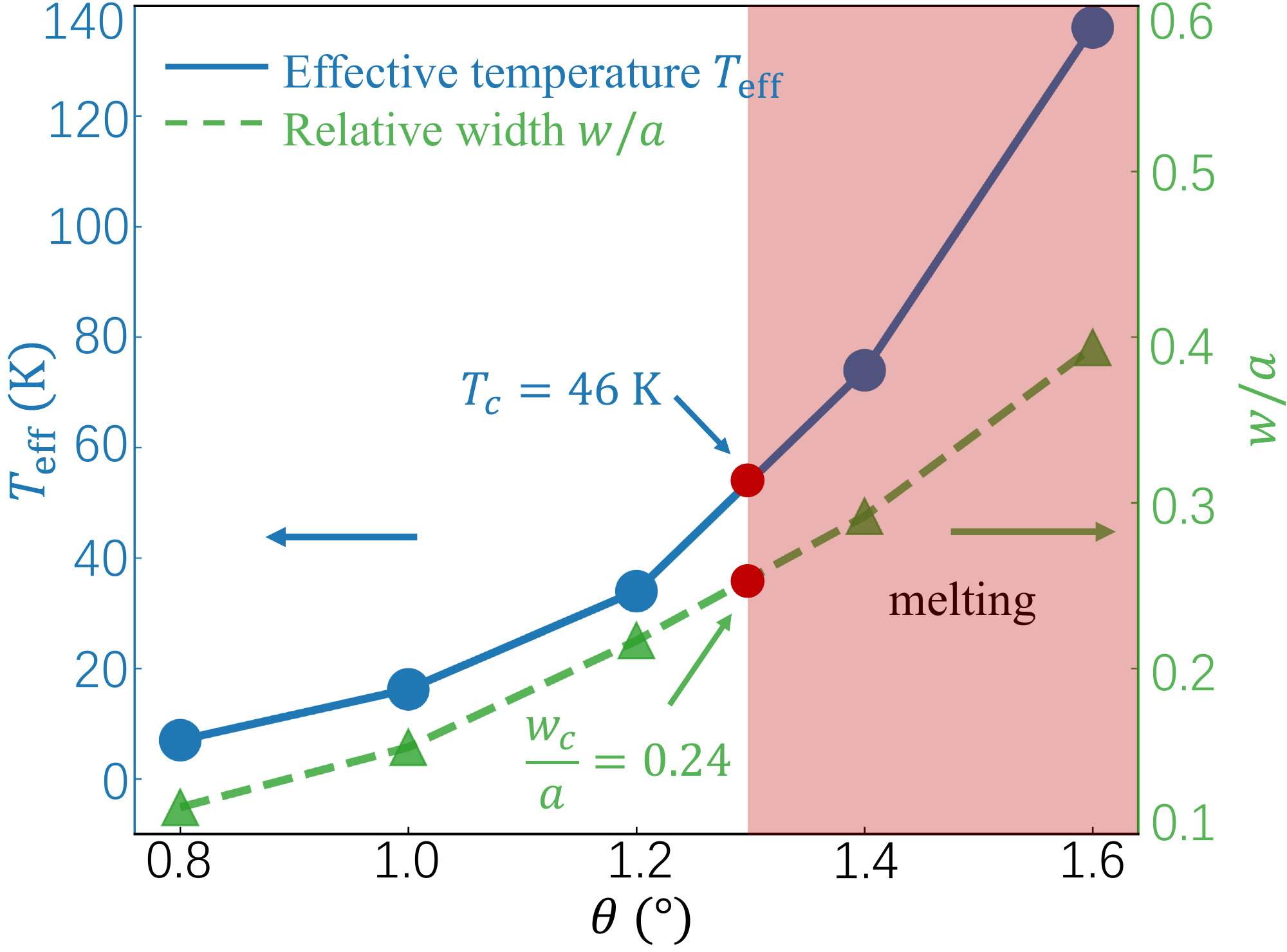}
	\caption{Results of Joule heating computed at several different twist angles for twisted bilayer TMDCs. 
    The pump duration is fixed as $T_0=10 \unit{ps}$ and the pump frequency is $\omega=\omega(q_M)$ for each twist angle.
    The pump field $E=E_c(\theta)$ is the threshold field to induce AFM order for each twist angle.
    The green line and data points show the relative nonthermal fluctuation $w/a$ of electrons in the Wigner lattice after pumping.
    The blue line and data points  are the corresponding effective temperature $T_{\mathrm{eff}}$. 
    The shaded region indicates the range of twist angle where the pump melts the Wigner crystal. 
   For all data points, we used the filling factor $\nu=1/10$, the pinning frequency $\omega_0=0.1 \unit{THz}$ and width $\gamma_0=0.1 \unit{THz}$, and the phononic damping rate $\gamma=0.05 \unit{THz}$, same as Figs.~2 and 3.
	\label{fig:heating_effect}}
\end{figure}

The absorbed energy density due to Joule heating from a pump pulse is 
\begin{equation}\label{heating}
	J_E \approx \frac{1}{2} \mathrm{Re}[\sigma(\omega)]E^2 T_0
	\,
\end{equation}
where $\omega$, $E$, and $T_0$ are the frequency, typical electric field, and duration of the pump pulse, respectively.
Summing over all the excited phonon modes, one obtains the local mean square fluctuation of an electron:
\begin{equation}\label{width_def}
	w^2=\frac{\hbar}{2m}\frac{1}{N}\sum_{q\alpha}\frac{2n_{q\alpha}+1}{\omega_{q\alpha}}\,
\end{equation}
where $n_{q\alpha}$ is the occupation number of the mode $q\alpha$.
Assuming that the absorbed energy is distributed in modes with frequencies near $\omega$ within a narrow window, the phonon occupation number $n_{q\alpha}=n_E \delta(\omega_{q\alpha}-\omega)$ is set by 
\begin{equation}\label{phonon_number}
	J_E=\int_{0}^{W}d\omega' \hbar \omega' g(\omega') n_E \delta(\omega'-\omega)
\end{equation}
where $W$ is the bandwidth of the electronic phonon. 
Note that with pinning, the bandwidth is slightly modified.
Considering that the typical pump frequency is on the order of the bandwidth $W$, we approximate the phononic density of states as $g(\omega')\propto\omega'$ (linear dispersion) and obtain the occupation as $n_E \approx J_E/n$ where $n$ is the electronic density. 
The mean square fluctuation is therefore
\begin{equation}\label{width}
	w^2 \approx \frac{\hbar}{2m}\int_{0}^{W}d\omega' g(\omega') \frac{2n_E \delta(\omega'-\omega)+1}{\omega'}
\,.
\end{equation}
We then compare the dimensionless number,  $w/a$, the fluctuation  of an electron normalized by the Wigner lattice constant $a$, to the thermal value $w_{\mathrm{thermal}}(T)/a$ to extract an effective temperature $T_{\mathrm{eff}}$.

In the monolayer case, the relative fluctuation $w/a$ reaches $\sim 31$ after $0.5~\unit{ps}$ of driving with the threshold electric field of $1.5\times 10^5 \unit{V/cm}$, far exceeding the melting threshold of the Wigner crystal. 
In contrast, in twisted bilayers at small twist angles, the required threshold field to induce the AFM order is much smaller.
The resulting $w/a$ is strongly suppressed (see Fig.~\ref{fig:heating_effect} for numerical results).
As the twist angle increases, the Berry curvature decreases, raising the critical field $E_c$ and increasing the heating effect.
The critical value of $w_c/a$ for melting is unclear.
We estimate it  from the melting point at the critical density $n_c = 5.7 \times 10^{12}~\unit{cm^{-2}}$ due to quantum  fluctuations found in a recent experiment \cite{xiang_imaging_2025}, which yields $w_c/a=0.24$  corresponding to an effective temperature of $46~\unit{K}$. 
In Fig.~\ref{fig:heating_effect}, we show the nonthermal fluctuation $w/a$ (green line) and the corresponding effective temperature (blue line)  after pumping for $T_0= 10~\unit{ps}$ computed for several different twist angles. There exists a wide range of twist angles for which $w/a$ is well below the melting threshold.

Note that after the stripe antiferromagnetic phase is favored during a THz pulse, the evolution to this phase occurs via phase-ordering dynamics~\cite{Bray1994PhaseOrdering,PhysRevX.10.021028}.
It is an interesting question whether sufficiently long-range order could be established before the injected heat is conducted from the phonons to the pseudospin sector, which warrants future study.
\\

\section{Berry phase term in the Lagrangian}
\label{app:berry_term}
To facilitate the reading, we show a brief derivation of the anomalous velocity term in the Lagrangian equation~(\ref{eqn:lagrangian_WC}) following Secs.~IV and V of Ref.~\cite{Xiao.2010}.
A semiclassical wave packet is expanded using single-band Bloch state $|\psi_{\mathbf{k}}\rangle=e^{i\mathbf{k}\cdot\mathbf{r}}|u_{\mathbf{k}}\rangle$ as
\begin{equation}\label{wave_packet}
	|W\rangle=\int d\mathbf{k} w(\mathbf{k}, t)|\psi_{\mathbf{k}}\rangle
\end{equation}
where $w(\mathbf{k}, t)$ is a sharp distribution around $\mathbf{k}_c$. The central position of the wave packet is 
$
	\mathbf{r}_c=\langle W|\hat{\mathbf{r}}|W\rangle
$.
Noting that 
$$
\langle\psi_{\mathbf{k'}}|\hat{\mathbf{r}}|\psi_{\mathbf{k}}\rangle = i\langle\psi_{\mathbf{k'}}|\partial_\mathbf{k}|\psi_{\mathbf{k}}\rangle - ie^{i(\mathbf{k}-\mathbf{k'})\cdot\mathbf{r}}\langle u_{\mathbf{k'}}|\partial_\mathbf{k}|u_{\mathbf{k}}\rangle,
$$ 
the central position is therefore
\begin{equation}\label{center_position}
	\mathbf{r}_c = -\partial_{\mathbf{k}}\ arg(w) \big|_{\mathbf{k}=\mathbf{k}_c} + \mathbf{A}(\mathbf{k}_c)
\end{equation}
where the Berry connection $\mathcal{A}$ is defined as
$
\mathbf{A}(\mathbf{k}) = i \left\langle u_{\mathbf{k}} \middle| \nabla_{\mathbf{k}} |u_{\mathbf{k}} \right\rangle.
$
In our convention, the semiclassical Lagrangian of the wave packet $|W\rangle$ is given by
\begin{equation}
L = \langle W |-i\partial_t + H | W \rangle.
\end{equation}
In the presence of external potential $V(\mathbf{r})$, the energy expectation value is $\langle W | H | W \rangle = \epsilon(\mathbf{k}_c)+V(\mathbf{r}_c)$ where $\epsilon(\mathbf{k})$ is the band dispersion. 
Using Eq.~(\ref{center_position}), the time-derivative term can be evaluated as
\begin{align}
\langle W | -i\partial_t | W \rangle 
&= \partial_t \arg w(\mathbf{k}_c,t) 
\notag\\
&= -\dot{\mathbf{k}}_c \cdot \left( \mathbf{A}(\mathbf{k}_c) - \mathbf{r}_c \right)
+ \frac{d}{dt} \arg w(\mathbf{k}_c,t).
\end{align}
Up to a total time derivative, the Lagrangian in Eq.~(1) takes the form
\begin{equation}
L =- \mathbf{k}_c \cdot \dot{\mathbf{r}}_c +\epsilon(\mathbf{k}_c) +V(\mathbf{r}_c)-  \mathbf{A}(\mathbf{k}_c) \cdot  \dot{\mathbf{k}}_c .
\label{Lagrangian}
\end{equation}
The last term is the geometric (Berry phase) contribution that gives rise to the anomalous velocity in the equations of motion.
\\


\begin{figure}
	\centering
	\includegraphics[width=0.7\linewidth]{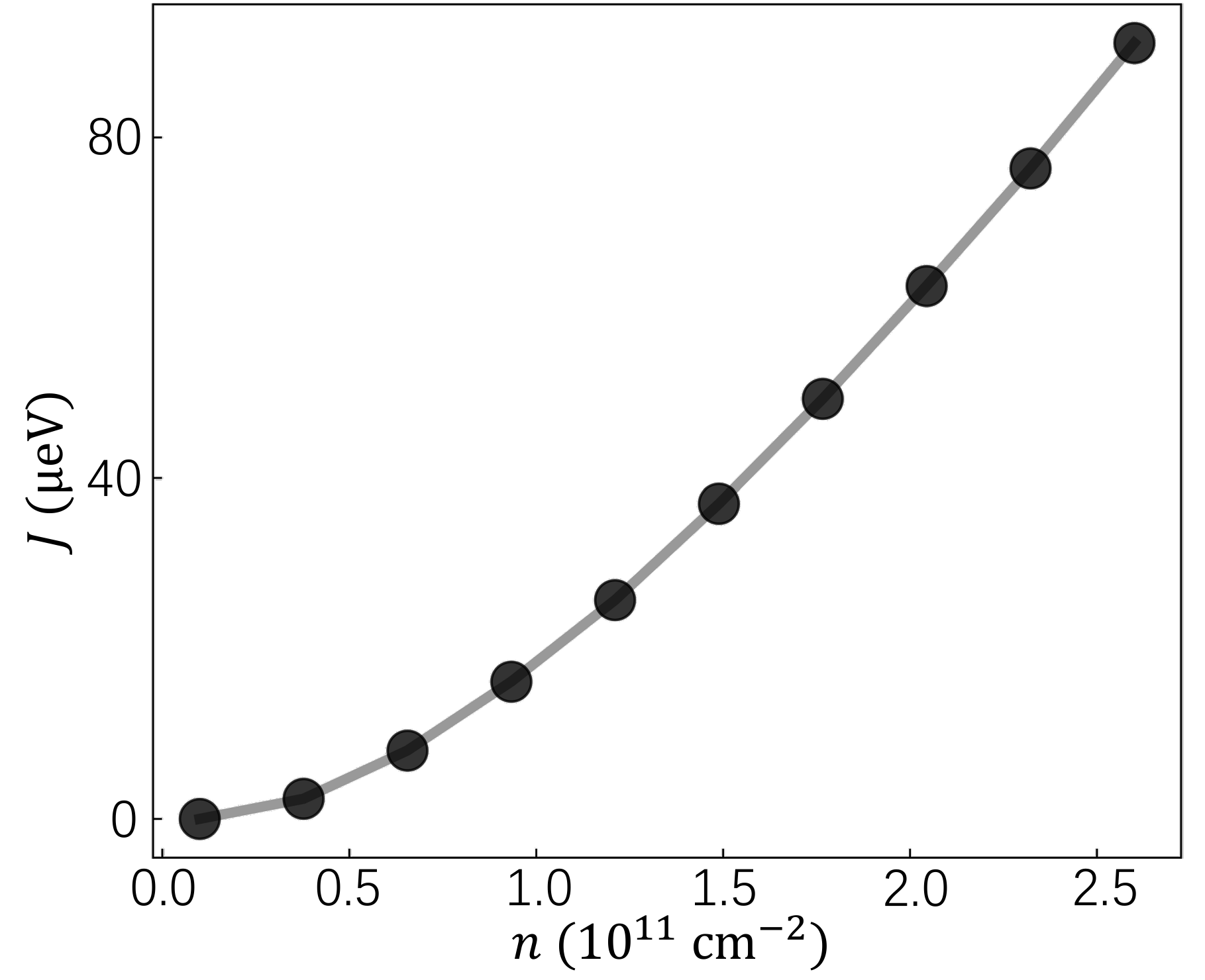}
	\caption{Carrier density dependence of exchange energy $J$. 
    We used effective mass $m=0.5m_e$ and environmental dielectric constant $\epsilon=7$ for monolayer TMDCs. For all chosen $n$, $r_s > 29.8$, exceeding the melting threshold~\cite{padhi_generalized_2021}.
	\label{fig:J_vs_n}}
\end{figure}

\section{Exchange energy}
\label{app:exchange}
In this Appendix, we estimate the order of magnitude of the coupling strength $J$ between adjacent pseudospins. More precise results can be obtained by computing the energy difference between the FM and AFM phases, as detailed in Table I of Ref.~\cite{drummond_phase_2009}.
The exchange coupling $J$ used in the main text is estimated from the exchange integral difference between parallel and antiparallel spin configurations of two adjacent electrons in the Wigner crystal, which is also consistent with other theoretical results from the literature~\cite{knorzer_wigner_2020}. 
The exchange integral $J_{ab}$ is 
$$
J_{ab}=\int d\mathbf{r}_1 d\mathbf{r}_2\psi_a^*(\mathbf{r}_1)\psi_b^*(\mathbf{r}_2)\frac{e^2}{4\pi\epsilon|\mathbf{r}_1-\mathbf{r}_2|}\psi_a(\mathbf{r}_2)\psi_b(\mathbf{r}_1)
\,.
$$
The local electronic wave functions are taken as
\begin{align}
\psi_a(r) &=\left( \dfrac{1}{2\pi w^{2}}\right) ^{1/2}\exp \left( -\dfrac{x^{2}+y^{2}}{4w^{2}}\right) e^{iKx}|\uparrow  \rangle
\,,
\notag\\
\psi_b(r) &=\left( \dfrac{1}{2\pi w^{2}}\right) ^{1/2}\exp \left( -\dfrac{\left( x-a\right) ^{2}-y^{2}}{4w^{2}}\right)
\begin{cases} 
e^{iKx}| \uparrow \rangle \\
 e^{iK'x}| \downarrow  \rangle
 \end{cases}
\end{align}
where the lattice constant $a$ and wave-packet width $w$ are determined by the carrier density $n$. 
The numerical estimate of the exchange energy $J$  in monolayer TMDCs at zero temperature is shown in \fig{fig:J_vs_n}.  It depends strongly on the density. 
In TMDC moir\'e bilayers, $J$ will be smaller due to larger effective masses and thus smaller wave-packet widths.
\\

\begin{figure}
	\centering
	\includegraphics[width=0.9\linewidth]{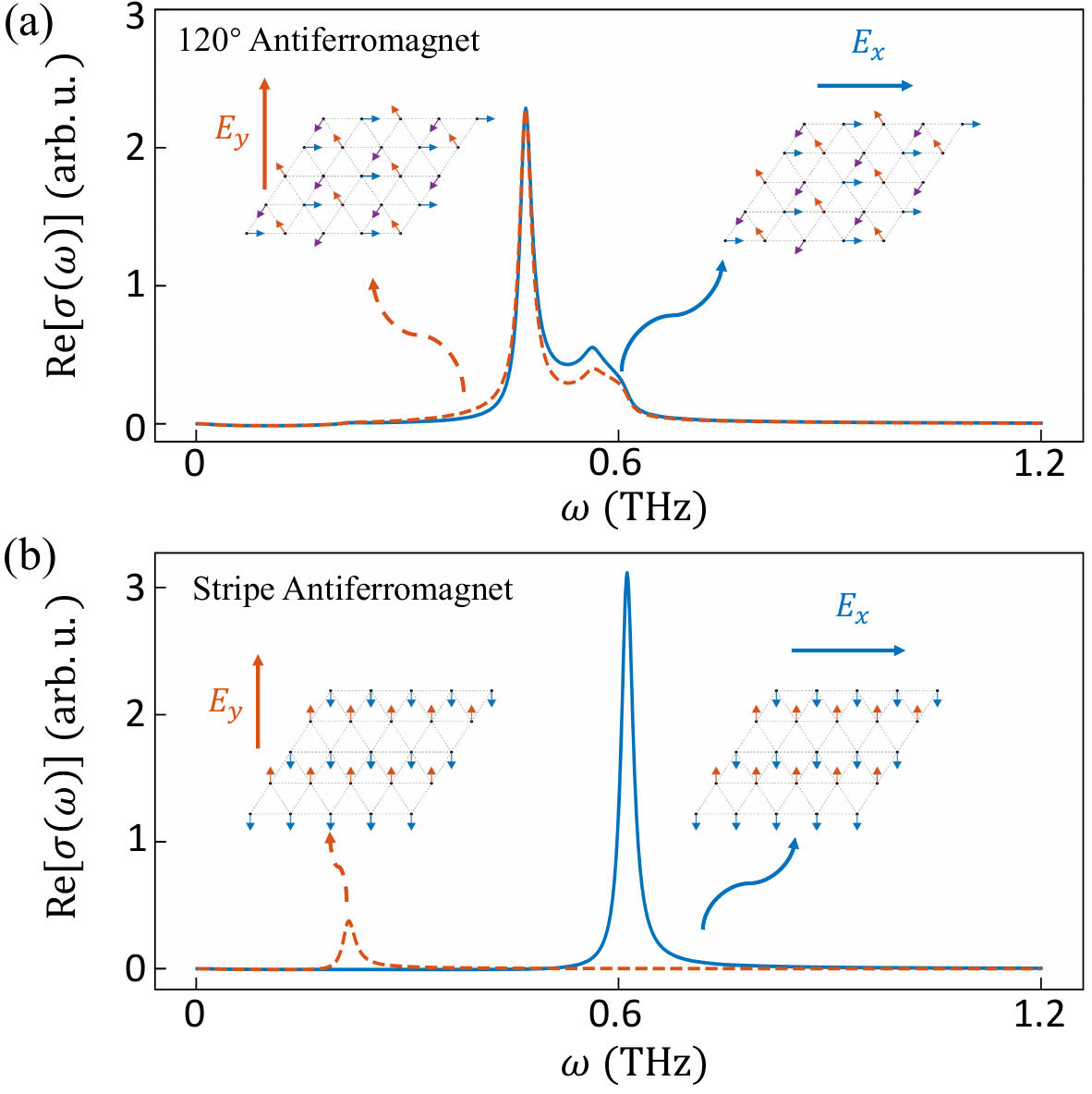}
	\caption{Dependence of the optical conductivity on the probe field polarization for (a) $120^{\circ}$ AFM phase  (b) and stripe AFM phase. 
    The (pinned) Drude peak is not shown.
    Blue and orange curves correspond to probe fields along the $x$ and $y$ directions, respectively (see schematic in the inset for relative directions of the probe fields and the pseudospin orders). 
    We used the carrier density $n=10^{11}\unit{{cm}^{-2}}$ and the effective mass $m=m_e$ motivated by the lowest two Chern bands in twisted bilayer MoSe$_2$~\cite{Wu.2019, regan_mott_2020, deng_nonmonotonic_2025} and the phononic damping rate $\gamma=0.02 \unit{THz}$. 
	\label{fig:direction_dependence}}
\end{figure}

\section{Distinguishing Pseudospin Orders in the optical conductivity}
\label{app:Distinguish}

In this Appendix, we show how to distinguish the peaks in the optical conductivity from different pseuduspin orders, and how to distinguish them from peaks due to other mechanisms.

The peaks from different types of pseudospin textures can be clearly distinguished from their structures as a function of frequency and probing direction.
As shown in \fig{fig:conductivity}(c), the spin liquid phase would result in a broad response that spans the whole bandwidth of the electronic phonons.
Because it has no (quasi-) long-range order, there are no sharp peaks.
Both of the two antiferromagnetic phases, the stripe AFM order and the 120$^\circ$ Néel AFM order, exhibit (quasi-) long-range order of the pseudospin, so that they generate sharp peaks in the optical conductivity [see \fig{fig:conductivity}(c)]. 
The best way to distinguish these two states is to look at the dependence on the polarization of the probe field.
The stripe AFM phase has a single ordering wave vector $\vec{q}$ that makes its optical response highly anisotropic.
As the polarization of the driving field changes from being perpendicular to  $\vec{q}$ to being parallel to $\vec{q}_0$, the location of the stripe AFM phase induced peak changes from that of the longitudinal electronic phonon frequency $\omega_{L,q_0}$ to that of the transverse phonon frequency $\omega_{T,q_0}$ [see \fig{fig:direction_dependence}(b)].
In contrast, the wave vectors of the $120^\circ$ AFM order are distributed at the six corners of the Brillouin zone\cite{hu_competing_2015}. Thus, the response of the 120$^\circ$ AFM state depends weakly on the probe polarization [see \fig{fig:direction_dependence}(a)]. 

It is also straightforward to distinguish the pseuduspin-order-induced peaks from those due to other mechanisms.
The most probable competing mechanisms that lead to THz peaks are the actual optical phonons of the hosting materials.
These phonons can be almost completely characterized experimentally from optical spectroscopy~\cite{Razzari.2025} and Raman spectroscopy~\cite{Zhang2015PhononRamanTMDC,
Liang2017LowFrequencyShearBreathing,
Cong2020Raman2D, Saito2016RamanTMDC, lee_anomalous_2010,lin_moire_2018}. 
First-principles calculations could also provide decent predictions for these phonons~\cite{Sohier2016FrohlichTMDC,Razzari.2025}.
Among the optical phonons of the materials, only the very few dipolar active (also called ``optically active'' or ``infrared active'') ones would contribute peaks in the THz optical spectroscopy that we proposed. 
As a benchmark for the magnitude and frequency scale of host dipole-active phonons, in monolayer WSe$_2$~\cite{Razzari.2025}, optics experiment found  one dipolar active phonon at $7.5 \unit{THz}$, which is the lowest-energy dipolar active phonon in this material.
In twisted bilayers, moir\'e folding may activate additional weak lattice-phononic features.
The frequencies and oscillator strengths of the host-material dipole-active optical phonons are set primarily by the lattice and therefore vary only weakly with carrier density.
In contrast, as we explained before,
the locations and spectra weights of pseudospin-order-induced THz peaks depend sensitively on the carrier density as certain power laws (see Appendix~\ref{app:moire}).
Therefore, they can be easily distinguished by looking at the doping level dependence.

Another competing mechanism for the peak is the pinned Drude peak by disorder.
One may use the power law dependence on the carrier density to distinguish it. 
As the carrier density is tuned, the pseudospin-order-induced peak frequency increases as $\omega_p =\omega_{q_0\alpha}\propto n^{3/4}$ (because it is locked to the electronic phonon frequency at the ordering wave vector) and its spectral weight increases as $W_p \propto \omega_p^2 \propto n^{3/2}$, while
the location of the Drude peak pinned by disorder
scales as $\omega_{\text{pinning}}\propto n^{3/8}$~\cite{chen_terahertz_2025} and its spectra weight scales as $W_{\text{pinning}}\propto n$.
Therefore, they follow completely different power laws.
Furthermore, because of the (quasi) long-range-ordered nature of the AFM pseudospin orders, they are expected to lead to peaks that are naturally very narrow, probably much narrower than the Drude peak.
The (pinned) Drude peak is always there and is easy to identify in an experiment~\cite{chen_terahertz_2025}. 
If the THz probe covers both the pinned Drude peak and the pseudospin-order-induced peak, the latter is most likely the higher-energy one (see \fig{fig:theta_dependence}).

The third competing mechanism is electronic interband transitions between minibands in TMDC moir\'e bilayers.
They typically give rise to a continuous spectrum rather than a sharp peak. 
Moreover, as the carrier density is varied, part of the spectra disappear or appear simply due to change of Pauli blocking, while the transition energies do not shift.
As discussed above, the carrier  density dependence of the pseudospin-order-induced THz peaks is completele different, thereby allowing for clear identification of them.
Even if interaction effects compress some of the interband transitions into excitonic excitations in the THz range, 
the locations of excitons have very little dependence on carrier density [see Fig.~1(d) of Ref.~\cite{chen_terahertz_2025} for the example of a high-energy exciton].
Moreover, the excitonic excitations exist primarily when integer numbers of moir\'e minibands are filled, while Wigner crystals exist only when a band is partially filled.
\\


\section{Spectral weight of the pseudospin-order-induced peak}
\label{app:spectral_weight}
In this Appendix, we estimate the spectral weight of the pseudospin-order-induced peak and compare it with those of optically active phonons from the host material itself~\cite{Razzari.2025}.
At the twist angle $\theta = 1.4^\circ$ and filling factor $\nu = 1/10$, the spectral weight of the pseudospin-order-induced peak is
$
W = 0.07 \frac{\pi n e^2}{2m},
$
as shown in Figs.~\ref{fig:theta_dependence} and \ref{fig:density_dependence}(c). In CGS units, $e^2 = 2.3 \times 10^{-19}~\unit{erg\cdot cm}$ and $m = m_e = 9.1 \times 10^{-28}~\unit{g}$, yielding
$
W = 0.72 \times 10^{19}~\unit{cm/s^2}
$.
For the actual phonons in the host material, we take a typical example in monolayer WSe$_2$~\cite{Razzari.2025}, where 
the lowest-energy dipole-active phonon was identified experimentally at $7.5~\unit{THz}$. From  Eq.~(1) of Ref.~\cite{Razzari.2025}, the equivalent in-plane three-dimensional  permittivity of this mode is
\begin{equation}
\epsilon_{\mathrm{eff}}(\nu) = 1 + \frac{1}{l} \frac{4 \epsilon_{\mathrm{env}} \nu_0 u_g}{\nu_0^2 - \nu^2 - i \nu \tau^{-1}},
\end{equation}
where the environmental permittivity is $\epsilon_{\mathrm{env}} = 1$, the phonon frequency is $\nu_0 = 7.47~\unit{THz}$, $l=0.65 \unit{nm}$ and  $u_g = 610~\unit{m/s}$.
The corresponding two-dimensional optical conductivity gives its spectral weight as
\begin{equation}
\int_0^{\infty}d\omega \mathrm{Re}\!\left[-i  \frac{\omega}{4\pi} \left(\epsilon_{\mathrm{eff}} - 1\right) l\right] = 0.89 \times 10^{19}~\unit{cm/s^2}
\end{equation}
which is comparable to the spectral weight of the pseudospin-order-induced peak.
Therefore, the pseudospin-order-induced peak is not weak compared to the phonons and would not be masked by the latter.  As explained in Appendix~\ref{app:Distinguish}, they can be distinguished by looking at the doping level dependence.
\\

\section{Nonperturbative results for the optical conductivity}
\label{app:self_energy}
In this Appendix, we expand \equa{eqn:solution} to infinite orders in the Berry curvature $\Omega$ and derive its self-energy correction to the  phonon propagator, so that the nonperturbative optical conductivity is obtained. Physically, this correction describes the renormalization of the phonon spectrum by the Berry curvature.

The renormalized phonon propagator in \equa{eqn:solution} is expanded as
\begin{equation}
G^\ast=
\frac{1}{G^{-1}+V}
=
\sum_{n=0}^{\infty}G(VG)^n,
\end{equation}
where $V=i\omega g\varphi$ is a "scattering potential" proportional to the Berry curvature. By resumming the momentum-conserving scattering processes, 
the  propagator at momentum $q$ becomes
\begin{equation}
\begin{aligned}
G_q^\ast
& \approx 
G_q
+\sum_{q'}
G_q V_{q-q'} G_{q'} V_{q'-q} G_q+\cdots
\\
&=
\frac{1}{G_q^{-1}-\Sigma(\omega,q)},
\end{aligned}
\end{equation}
where the self-energy is 
\begin{equation}\label{eqn:sigma_SI}
\Sigma(\omega,q)
=
\left(
\omega\frac{m\Omega}{\hbar}
\right)^2
\sum_{q'}
\frac{
\omega_q^2\omega_{q'}^2\chi_{q-q'}
}{
\omega^2-\omega_{q'}^2+i\gamma\omega
}
\end{equation}
to second order in the Berry curvature.
This corresponds to the second-order Born approximation.
With this self-energy, the optical conductivity in \equa{sigma} is modified to the nonperturbative version:
\begin{equation}\label{sigma_Appendix}
\sigma(\omega)=ne^2 m \Omega^2 
\sum_{\alpha q} 
\frac{i\omega \omega_{q\alpha}^2  |v^y_{\alpha,q}|^2\chi_q}{\omega^2+i \gamma \omega -\omega_{\alpha,q}^2 - \Sigma(\omega, q)} 
\,.
\end{equation}
Equation~\eqref{sigma_Appendix} makes it explicit that beyond our lowest-order result in \equa{sigma}, the higher-order terms are controlled by  the self-energy as the  expansion parameter.

In \equa{eqn:sigma_SI}, $\omega m\Omega/\hbar$ is a dimensionless expansion parameter, which is $\sim 0.1$ in  Moire bilayer TMDCs with typical Moire periods (see Appendix~\ref{app:berry_curvature}). 
Because the pseudospin correlation function satisfies the normalization sum rule $\sum_q \chi_q = 1$  over the Brillouin zone, the self-energy is bounded by
\begin{equation}
\left|\Sigma(\omega,q)\right|
\lesssim
\left(
\omega\frac{m\Omega}{\hbar}
\right)^2
W^2
\ll W^2
\end{equation}
where $W$ is the bandwidth of the electronic phonons. 
Therefore, relative to typical phonons within the band, the Berry-curvature-induced frequency shift is small even for typical TMDC moire bilayers, 
justifying the lowest-order result used in the main text.
\\

\section{Fluctuation effects on the light-induced pseudospin order}
\label{app:fluctuation}
The light-induced stripe AFM state is an Ising type of order that spontaneously breaks the translational symmetry and the $Z_2$ symmetry of $S_z \rightarrow - S_z$. 
If the favored wave vector $q_c$ is commensurate with the underlying electronic lattice, the broken symmetry is discrete and this phase will be  stable below a nonzero critical temperature set roughly by $T_c\sim C_{q_c}$.
For the driving frequency $\omega = \omega(q_{M})$ and at the driving field $E=3 \times10^{5} \unit{V/cm}$, this temperature  is estimated to be $T_c \sim 1 \unit{K}$ for TMDC monolayers. 
In TMDC moir\'e structures where the typical Berry curvature of the lowest electronic bands is $\sim 100$ times larger~\cite{Wu.2019} because of much smaller Brillouin zones, the critical temperature is  boosted to  $T_c \sim 7 \unit{K}$ for an even weaker driving field $E=6\times 10^3 \unit{V/cm}$.
If the favored wave vector is incommensurate and if the continuous rotational symmetry around $S_z$ is also spontaneously broken, the fluctuations are stronger and  remain to be addressed by future research.

	\bibliography{ref_spin}
    
\end{document}